\documentclass[aps,pra,twocolumn,showpacs,superscriptaddress,longbibliography]{revtex4-1}
\usepackage{graphicx}

\usepackage{amsmath}

\usepackage{amssymb}

\usepackage{epstopdf}

\usepackage{amsfonts}

\usepackage{dcolumn}

\usepackage{multirow}

\usepackage{multirow}
\usepackage{physics}
\usepackage{CJK}
\usepackage[utf8]{inputenc}
\usepackage{esint}
\usepackage{float}
\usepackage{breakurl}
\usepackage{hyperref}
\hypersetup{colorlinks,citecolor=blue}

\makeatletter
\@ifundefined{textcolor}{}
{
 \definecolor{BLACK}{gray}{0}
 \definecolor{WHITE}{gray}{1}
 \definecolor{RED}{rgb}{1,0,0}
 \definecolor{GREEN}{rgb}{0,1,0}
 \definecolor{BLUE}{rgb}{0,0,1}
 \definecolor{CYAN}{cmyk}{1,0,0,0}
 \definecolor{MAGENTA}{cmyk}{0,1,0,0}
 \definecolor{YELLOW}{cmyk}{0,0,1,0}
}

\usepackage{subfigure}\usepackage{epsfig}\usepackage{amsfonts}\usepackage{mathrsfs}
\usepackage[toc,page,title,titletoc,header]{appendix}
\usepackage{amsmath}

\makeatother
\begin{document}

\title{Laser Control of Singlet-Pairing Process in an Ultracold Spinor Mixture}

\author{Jianwen Jie}

\affiliation{Shenzhen Institute for Quantum Science and Engineering,
Southern University of Science and Technology, Shenzhen, 518055, China}

\affiliation{Department of Physics, Renmin University of China, Beijing, 100872,
China}

\author{Yonghong Yu}

\affiliation{Department of Physics, Renmin University of China, Beijing, 100872,
China}

\author{Dajun Wang}

\affiliation{Department of Physics, The Chinese University of Hong Kong, Hong Kong SAR, China}

\author{Peng Zhang}
\email{pengzhang@ruc.edu.cn}

\affiliation{Department of Physics, Renmin University of China, Beijing, 100872,
China}

\affiliation{Beijing Computational Science Research Center, Beijing, 100084, China}

\affiliation{Beijing Key Laboratory of Opto-electronic Functional Materials \&
Micro-nano Devices (Renmin University of China)}

\date{today}

\begin{abstract}

In the mixture of ultracold spin-1 atoms of two different species A and B (e.g., $^{23}$Na (A) and $^{87}$Rb (B)),
 inter-species singlet-pairing process ${\rm A}_{+1}+{\rm B}_{-1}\rightleftharpoons {\rm A}_{-1}+{\rm B}_{+1}$
can be induced by the spin-dependent inter-atomic interaction, where subscript $\pm 1$ denotes the magnetic quantum number. Nevertheless, one cannot isolate this process from other spin-changing processes, which are usually much stronger, by tuning the bias real magnetic field. As a result, it is difficult to clearly observe singlet-pairing process and precisely measure the corresponding interaction strength.
In this work we propose to control the  singlet-pairing process via combining the real magnetic field and a laser-induced species-dependent synthetic magnetic field. With our approach one can significantly enhance this process  and simultaneously suppress all other spin-changing processes. We illustrate our approach for both a confined two-atom system and a binary mixture of spinor Bose-Einstein condensates. Our control scheme is helpful for the precise measurement of the weakly singlet-pairing interaction strength and  the entanglement generation of two different atoms.
\end{abstract}
\maketitle

\section{Introduction}
In last few decades, spinor Bose-Einstein condensates (BECs) was one of the most inspiring workhorses for studying diverse physics including spin textures, topological excitation and non-equilibrium quantum dynamics  \cite{PR2012_Spinor_BEC:Ueda,RMP2013_Spinor_BEC:Ueda}. In recent years, the vitality expansion of spinor BECs is well done by experimentally realizing of plentiful dramatical physical scenes, including SU(1,1) interferometer  \cite{PRLSmerzi2015,PRL2016Linnemann,PRL2017Szigeti,PRA2018_SU11:P.D.Lett,PRAJie2019}, quantum synchronization  \cite{Roulet2018PRL,laskar2020PRL}, gauge invariance  \cite{Mil1128}, entanglement generation  \cite{ScienceYouLi2017,PANS2018Zou} and dynamical quantum phase transitions  \cite{PRL2020Duan}.

In a single-species spinor  BEC, e.g., the BEC of spin-1 $^{87}$Rb   \cite{Rb2001PRL} or $^{23}$Na   \cite{Na1998PRL} atoms,  the spin-dependent inter-atomic interaction can induce a two-body spin-mixing process ${\rm A}_{0}+{\rm A}_{0}\rightleftharpoons {\rm A}_{+1}+{\rm A}_{-1}$, where A denotes the atomic species (e.g., $^{87}$Rb  or $^{23}$Na) and the subscript $0,\pm 1$ denotes the magnetic quantum number of atomic spin. This process can induce fruitful spin dynamics which were successfully observed  \cite{chang2005coherent,PRA2005Sengstock,PRL2007Lett} and can be used for the generation of spin squeezing, entanglement state as well as quantum metrology gain  \cite{PRL2007Vengalattore,ma2011quantum,hamley2012spin,PRLSmerzi2015,PRL2016Linnemann,PRL2017Szigeti,RMP2018:Smerzi,PRAJie2019,PRA2019Qimin}.

Furthermore, the binary mixture of spin-1 bosonic atoms has also been experimentally realized with several atomic combinations  \cite{PRL2015Dajun,Lin2020,Mil1128,fang2020collisional}. In such  two-species system the spin-dependent inter-species interaction can induce various types of spin-changing processes for two different atoms, i.e., the spin-mixing processes ${\rm A}_{0}+{\rm B}_{0}\rightleftharpoons {\rm A}_{\pm1}+{\rm B}_{\mp1}$ as well as the spin-exchanging processes ${\rm A}_{0}+{\rm B}_{\pm 1}\rightleftharpoons {\rm A}_{\pm 1}+{\rm B}_{0}$ and ${\rm A}_{+1}+{\rm B}_{-1}\rightleftharpoons {\rm A}_{-1}+{\rm B}_{+1}$, where A and B denote the atomic species and the subscript $0,\pm 1$ denotes the atomic magnetic quantum number, as above. These processes can induce coherent heteronuclear spin dynamics   \cite{Xu2012PRA,Chen2018PRA} and significantly influence the quantum phases of the two-species spin-1 BEC  \cite{Luo2007PRA,XuPRA2009,ShiPRA2010,Zhang2011PRA,Li_2017,XuPRA2010}. In the experiments one can control  most of the above spin-changing processes via the bias magnetic field. Explicitly, to enhance one specific process, one can just tune the magnetic field to a particular value so that the  total Zeeman energy of the two atoms before  this process is close to the one after this process, i.e., the initial and finial two-atom spin state of this process is near ``resonant" with each other. Using this technique Li {\it et al.} successfully observed the spin-exchanging processes ${\rm A}_{0}+{\rm B}_{\pm 1}\rightleftharpoons {\rm A}_{\pm 1}+{\rm B}_{0}$ in the mixture of ultracold $^{87}$Rb and $^{23}$Na atoms  \cite{PRL2015Dajun}.

The process ${\rm A}_{+1}+{\rm B}_{-1} \rightleftharpoons {\rm A}_{-1}+{\rm B}_{+1}$ is also called as the ``singlet-paring process" (or ``quadrupole exchange process" \cite{fang2020collisional}).
In this process
magnetic quantum number of each atom can be changed by $\pm2$, while in
all other processes $\langle1\rangle-\langle6\rangle$ the single-atom
magnetic quantum number can only be changed by $\pm1$. However, this interesting process cannot be controlled via the the above  approach. This can be explained as following. As shown below (Sec. III. A), for the singlet-paring process the above Zeeman-energy ``resonant" condition can be satisfied only when the bias magnetic field is zero. Nevertheless, in this case such ``resonant" condition is also satisfied for all other spin-changing processes. As a result, the singlet-paring process would be mixed with other processes and thus cannot be clearly detected. 
Moreover, the singlet-paring process is very weak. For instance, 
its strength is only 0.8\% compare to the strength of other spin-exchange processes for the $^{87}$Rb-$^{23}$Na mixture.
Due to these facts, it is difficult to clearly observe spinglet-pairing process and precisely measure the corresponding interaction strength \cite{fang2020collisional}.


In this work we propose an approach to effectively controlling the singlet-pairing process. Our basic idea is to apply both the real magnetic field and the synthetic magnetic field (SMF) induced by the vector light shift of a circular-polarized laser beam{\color{black}, which was found in atom physics in 1970s \cite{PRA1972} and has been used experimentally in the field of
ultracold spinor gases
  \cite{PRL2015Dajun,Goldman_2014}.} In this case the total ``effective magnetic field" experienced by each atom would be the summation of the real magnetic field and the SMF.
An important property of the SMF is that it is species-dependent. As a result, in the presence of the SMF the total ``effective magnetic field" experienced by  atoms of different species would be different, and can be {\it independently} controlled, as illustrated in the experiment of Ref.~\cite{PRL2015Dajun}. Therefore, one can tune the system to some points where the singlet-pairing process is energetically resonant,  while the other spin-changing processes are far-off resonant. Around this point the  singlet-pairing process can be significantly enhanced and isolated from other processes.

In  previous experiments of ultracold $^{87}$Rb and $^{23}$Na atoms   \cite{PRL2015Dajun,Lin2020} the SMF has been illustrated for the  manipulation of the spin-exchange process ${\rm A}_{0}+{\rm B}_{-1}\rightleftharpoons {\rm A}_{-1}+{\rm B}_{0}$ and the spin-mixing process ${\rm A}_{0}+{\rm B}_{0}\rightleftharpoons {\rm A}_{+1}+{\rm B}_{-1}$. Nevertheless, in the absence of the SMF,
these two processes can still be enhanced via real magnetic fields, with the ``resonant" method mentioned above. The experiments in Refs.   \cite{PRL2015Dajun,Lin2020} show that in the presence of the SMF the values of the real magnetic field required to enhance these two processes are shifted. For our case, as shown above, one cannot enhance the singlet-pairing process and simultaneously isolate it from other processes only with the real magnetic field. Thus the application of the SMF is necessary.


In the following sections  we take the mixture of ultracold $^{87}$Rb and $^{23}$Na atoms as an example, and illustrate our approach for both a confined two-atom system and the binary mixture of BECs of $^{87}$Rb and $^{23}$Na atoms. Our approach can be used for the observation and manipulation of the singlet-paring process, the precise measurement of the corresponding interaction intensity,  as well as the entanglement generation of two different spin-1 ultracold atoms.

The remainder of this article is organized as follows.  In Sec.~\ref{secII} we introduce the inter-atomic interactions and related spin-changing processes of the mixture of ultracold $^{87}$Rb and $^{23}$Na atoms. In Sec.~\ref{secIII} our proposal for the manipulation of singlet-pairing process is introduced. In Secs.~\ref{secIV} and \ref{secV} we further illustrate our proposal for a confined two-atom system and a binary mixture of BECs, respectively.  A  summary  for  our  results  and  some  discussions
are  given  in  Sec.~\ref{secVI}.  In  the  appendix  we  present some  details  of  our  calculation.

\section{spin-changing scattering process between ultracold spin-1 bosons}\label{secII}

We consider the mixture of ultracold spin-1 $^{87}$Rb  and $^{23}$Na atoms
at low magnetic field. In this system the inter-atomic interaction seriously depends on the atomic species. Explicitly, when the two atoms are of the same species $j$ ($j=$Rb and Na for $^{87}$Rb
and $^{23}$Na, respectively) the interaction is given by   \cite{PR2012_Spinor_BEC:Ueda,RMP2013_Spinor_BEC:Ueda} 
\begin{equation}
\hat{U}_{j}({\bf r})=\left(\alpha_{j}+\beta_{j}\hat{\textbf{F}}_{1}\hat{\cdot\textbf{F}}_{2}\right)\delta({\bf r}),\label{uj}
\end{equation}
when the two atoms are of different species the interaction is given by  \cite{PRL2015Dajun}
\begin{equation}
\hat{U}_{{\rm Rb-Na}}({\bf r})=\left(\alpha+\beta\hat{\textbf{F}}_{1}\cdot\hat{\textbf{F}_{2}}+\gamma{\cal \hat{P}}_{0}\right)\delta({\bf r}).\label{uab}
\end{equation}
Here ${\bf r}$ is the inter-atomic relative coordinate, $\hat{{\bf F}}_{1}$ and $\hat{{\bf F}}_{2}$ are the respective spin operators of the two atoms,
and $\hat{{\cal P}}_{0}$ is
the projection operator for the two-body hyperfine state corresponding
to total spin $\hat{\textbf{F}}_{\rm tot}\equiv\hat{\textbf{F}}_{1}+\hat{\textbf{F}}_{2}=0$. The interaction intensities $(\alpha_{j},\beta_{j},\alpha,\beta,\gamma)$
($j=$Rb, Na) are given by 
\begin{eqnarray}
\alpha_{j} & = & \left(\frac{4\pi\hbar^{2}}{3M_{j}}\right){\left(a_{j}^{(0)}+2a_{j}^{(2)}\right)};\label{lj}\\
\beta_{j} & = & \left(\frac{4\pi\hbar^{2}}{3M_{j}}\right){\left(a_{j}^{(2)}-a_{j}^{(0)}\right)};\label{ej}
\end{eqnarray}
and 
\begin{eqnarray}
\alpha & = & \left(\frac{\pi\hbar^{2}}{\mu}\right){\left(a_{{\rm Rb-Na}}^{(0)}+2a_{{\rm Rb-Na}}^{(2)}\right)};\label{alpha}\\
\beta & = & \left(\frac{\pi\hbar^{2}}{\mu}\right){\left(a_{{\rm Rb-Na}}^{(2)}-a_{{\rm Rb-Na}}^{(1)}\right)};\label{beta}\\
\gamma & = & \left(\frac{\pi\hbar^{2}}{\mu}\right){\left(2a_{{\rm Rb-Na}}^{(0)}-3a_{{\rm Rb-Na}}^{(1)}+a_{{\rm Rb-Na}}^{(2)}\right)}.\label{gamma}
\end{eqnarray}
Here $M_{j}$ and $a_{j}^{(F_{\rm tot})}$ ($j=$Rb, Na; $F_{{\rm tot}}=0,2$)
are the mass and $s$-wave scattering length of a single atom
of type $j$, respectively; {\color{black}while $\mu=\frac{M_{\rm Rb}M_{\rm Na}}{M_{\rm Rb}+M_{\rm Na}}$}, and $a_{{\rm Rb-Na}}^{(F_{{\rm tot}})}$
($F_{{\rm tot}}=0,1,2$) are the reduced mass and $s$-wave scattering
length of two atoms of different species with total spin $F_{{\rm tot}}$,
respectively. Previous measurements show that $(\alpha_{{\rm Rb}}, \beta_{{\rm Rb}})=\frac{4\pi\hbar^{2}a_{0}}{M_{{\rm Rb}}}(100.9, -0.47)$  \cite{RMP2013_Spinor_BEC:Ueda},
$(\alpha_{{\rm Na}}, \beta_{{\rm Na}})=\frac{4\pi\hbar^{2}a_{0}}{M_{{\rm Na}}}(52.66,1.88)$ \cite{PRA2011Tiemann}. The theoretical calculations show that $(\alpha, \beta)=\frac{2\pi\hbar^{2}a_{0}}{\mu}(78.9, -2.5)$  and $\gamma=(0.06)\frac{2\pi\hbar^{2}a_{0}}{\mu}$ \cite{NaRbSL,PRL2015Dajun},
with $a_{0}$ being the Bohr radius.

Since both $^{87}$Rb  and $^{23}$Na atoms are considered in the $F=1$ hyperfine manifold, each atom has three
hyperfine states corresponding to magnetic quantum number
$m=0,\pm1$. The above two-body interactions can induce the following seven  
spin-changing scattering processes:
\begin{eqnarray}
\left\langle 1\right\rangle  & \beta_{{\rm Rb}}: & {\rm Rb}_{-1}+{\rm Rb}_{+1}\rightleftharpoons{\rm Rb}_{0}+{\rm Rb}_{0};\nonumber\\
\left\langle 2\right\rangle  & \beta_{{\rm Na}}: & {\rm Na}_{-1}+{\rm Na}_{+1}\rightleftharpoons{\rm Na}_{0}+{\rm Na}_{0};\nonumber\\
\left\langle 3\right\rangle  & \beta-\gamma/3: & {\rm Rb}_{-1}+{\rm Na}_{+1}\rightleftharpoons{\rm Rb}_{0}+{\rm Na}_{0};\nonumber\\
\left\langle 4\right\rangle  & \beta-\gamma/3: & {\rm Rb}_{+1}+{\rm Na}_{-1}\rightleftharpoons{\rm Rb}_{0}+{\rm Na}_{0};\label{sps}\\
\left\langle 5\right\rangle  & \beta: & {\rm Rb}_{0}+{\rm Na}_{+1}\rightleftharpoons{\rm Rb}_{+1}+{\rm Na}_{0};\nonumber\\
\left\langle 6\right\rangle  & \beta: & {\rm Rb}_{0}+{\rm Na}_{-1}\rightleftharpoons{\rm Rb}_{-1}+{\rm Na}_{0};\nonumber\\
\left\langle 7\right\rangle  & \gamma/3:&  {\rm Rb}_{-1}+{\rm Na}_{+1}\rightleftharpoons {\rm Rb}_{+1}+{\rm Na}_{-1}\nonumber\\
  && ({\rm singlet}\ {\rm pairing}).\nonumber
\end{eqnarray}
In Eq. (\ref{sps}) we also show the corresponding interaction intensity before each reaction
equation, and the subscripts of $\pm1,0$ denote the magnetic quantum
number $m$ of each atom. For instance, in the process $\langle1\rangle$
the magnetic quantum numbers $(m_{1},m_{2})$ of the two  $^{87}$Rb atoms can
be changed from $(-1,+1)$ to $(0,0)$ and vice versa. The processes
$\langle1,2\rangle$ and $\langle3\rangle-\langle7\rangle$ are intra-species and inter-species spin-changing collisions, respectively. 

The above process $\langle7\rangle$  is the singlet-pairing process  \cite{fang2020collisional}, as mentioned above. In this process the
magnetic quantum number of each atom can be changed by $\pm2$, while in
all other processes $\langle1\rangle-\langle6\rangle$ the single-atom
magnetic quantum number can only be changed by $\pm1$. In the following section we
show our approach for the laser control of this process.
\section{Control of singlet-pairing process via laser-induced SMF}\label{secIII} 

In this work consider the cases with low magnetic field (less than four Gauss).  Under this condition no magnetic Feshbach resonance \cite{FR2010RMP,NaRbSL} for our system has been discovered, and thus
the interaction intensities $(\alpha_{j},\beta_{j},\alpha,\beta,\gamma)$
($j=$Rb, Na) cannot be changed via the magnetic field. Nevertheless,  one can still efficiently control the 
 spin-changing processes by changing the detuning between 
 the two-atom Zeeman-energies before and after each process.
This detuning can be denoted as $\Delta_{\left\langle l\right\rangle }$ for
the process $\left\langle l\right\rangle $ ($l=1,2,...,7$). For
instance, for the singlet-pairing process $\left\langle 7\right\rangle $ we have
\begin{equation}
\Delta_{\left\langle 7\right\rangle }=\left(E_{-1}^{({\rm Rb})}+E_{+1}^{({\rm Na})}\right)-\left(E_{+1}^{({\rm Rb})}+E_{-1}^{({\rm Na})}\right),\label{d7}
\end{equation}
with $E_{m}^{(j)}$ ($j=$Rb, Na; $m=0,\pm1$) being the free energy
of a $j$-atom with magnetic quantum number $m$. 
For our weakly-interacting systems, the effect of the spin-changing process
$\left\langle l\right\rangle $ ($l=1,2,...,7$) is usually significant
when $\Delta_{\left\langle l\right\rangle }=0,$ i.e., when the hypferfine
states before and after the scattering are resonant with each other.
Accordingly, the effect of process $\left\langle l\right\rangle $
is weak when the detuning $\Delta_{\left\langle l\right\rangle }$
is far away from zero.

Our purpose of this work is to enhance the effect of the singlet-pairing
process $\left\langle 7\right\rangle $ and simultaneously suppress
the effect of other processes. According to the above discussion,
we can realize this by tuning the detuning $\Delta_{\left\langle 7\right\rangle }$
to zero while keeping the detunings for other processes to finite,
i.e., 
\begin{eqnarray}
&&\Delta_{\left\langle 7\right\rangle }=0;\label{cc1}\\
&&\Delta_{\left\langle l\right\rangle }\neq0\ {\rm for}\ l=1,...,6.\label{cc}
\end{eqnarray}
{\color{black}For realistic systems the above condition can be expressed more explicitly as
\begin{eqnarray}
\abs{\frac{\Delta_{\left\langle 7\right\rangle }}{g_{\left\langle 7\right\rangle}}}&\ll& 1;\\
\abs{\frac{\Delta_{\left\langle l\right\rangle }}{g_{\left\langle l\right\rangle}}}
&\gtrsim& 1,\hspace{0.2cm}
 {\rm for}\ l=1,...,6,
\end{eqnarray}
where $g_{\left\langle l\right\rangle}$  ($l=1,...,7$) is the characteristic interaction intensity  for the process $\left\langle l\right\rangle$ \cite{resonant_conditions}.}

\subsection{The case only with the real magnetic field}

We first consider the case without laser-induced species-dependent SMF. In this case the free energies $E_{m}^{(j)}$ ($j=$Rb, Na;
$m=0,\pm1$) as well as the detunings $\Delta_{\left\langle l\right\rangle }$
($l=1,2,...,7$) can only be tuned via the real magnetic field $B{\bf e}_z$, with ${\bf e}_z$ being the unit vector along the $z$-direction. Explicitly, we have 
\begin{equation}
E_{m}^{(j)}=-p^{(j)}mB+q^{(j)}m^{2}B^{2}\ \ \ (j={\rm Rb},\ {\rm Na}).\label{emj}
\end{equation}
Here the first and second terms are the linear and
quadratic Zeeman effects, respectively, with $p^{(j)}$ and $q^{(j)}$
being the corresponding coefficients. Explicitly, we have $(p^{({\rm Rb})},\ p^{({\rm Na})})/h=(702369,702023)$ (Hz/G)
and $(q^{({\rm Rb})},\ q^{({\rm Na})})/h=(72,277)$ (Hz/G$^{2}$) \cite{Rbdata,Nadata}.

According to Eq. (\ref{emj}) and Eq. (\ref{d7}), the
detuning $\Delta_{\left\langle 7\right\rangle }$ for the singlet-pairing
process $\left\langle 7\right\rangle $ is 
\begin{equation}
\Delta_{\left\langle 7\right\rangle }=2\left(p^{({\rm Rb})}-p^{({\rm Na})}\right)B.\label{d72}
\end{equation}
Since $p^{({\rm Rb})}\neq p^{({\rm Na})}$, Eq. (\ref{d72}) yields
that $\Delta_{\left\langle 7\right\rangle }$ is zero only when $B=0$.
However, according to Eq. (\ref{emj}), in
this case the detunings $\Delta_{\left\langle 1,..,6\right\rangle }$
of other spin-changing processes are all zero. {\color{black}Therefore, 
the conditions (\ref{cc1}) and
(\ref{cc}) 
cannot be simultaneously
satisfied with only a real magnetic field and no synthetic field.}

\subsection{Our proposal }

\begin{figure}
\includegraphics[width=5.7cm]{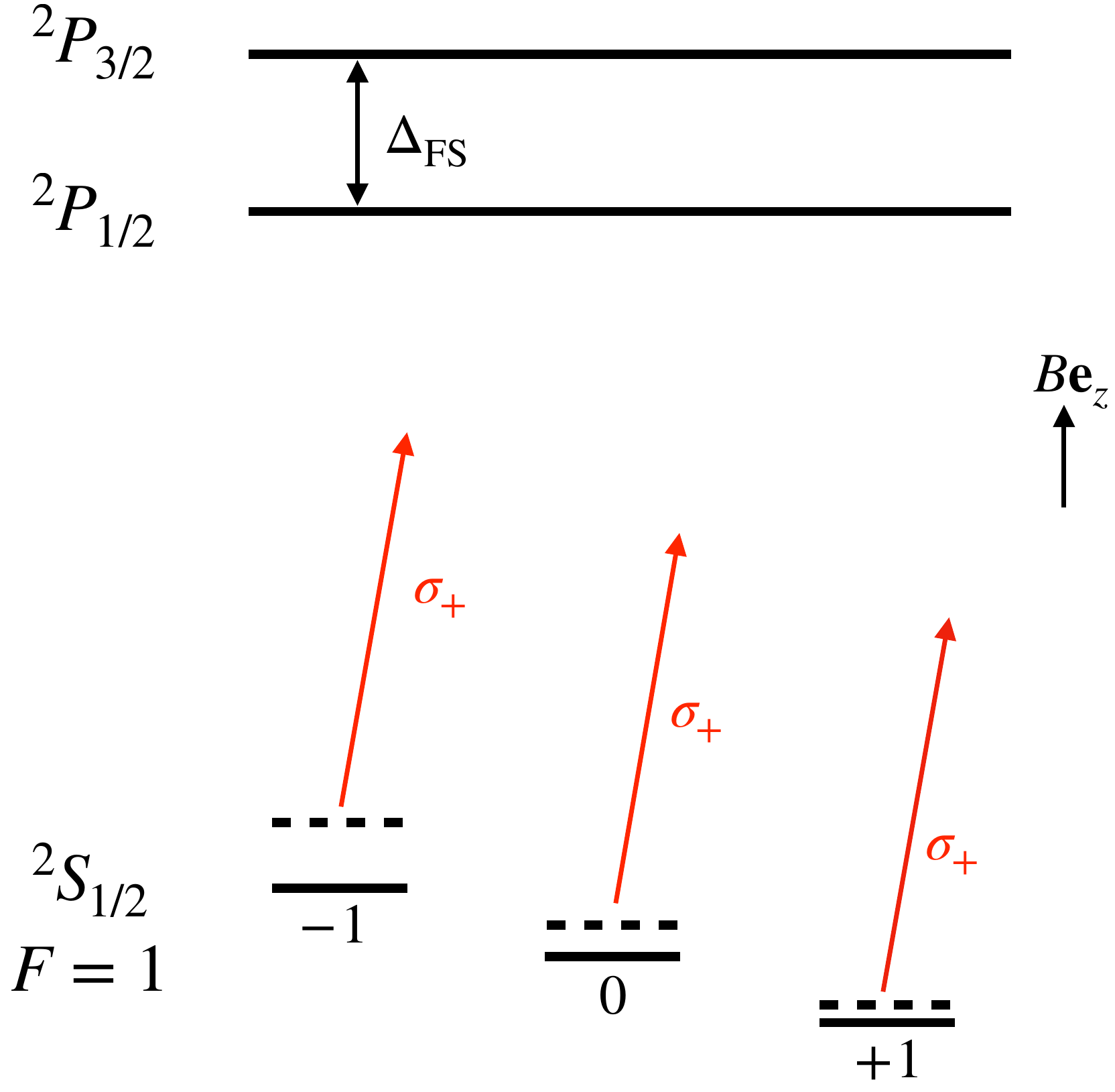} 
\caption{(Color online) Schematic diagram for the laser-induced SMF for a spin-1 ultracold alkaline atom, i.e., the result in Eq. (\ref{demj}). The solid {\color{black}black} lines are the Zeeman levels of the $^2$S$_{1/2}$  states with $F=1$ for the case without the laser beam, which is totally determined by the real magnetic field $B{\bf e}_z$. The dashed {\color{black}black} lines are the levels with both the real magnetic field and a $\sigma_+$-polarized laser beam. $\Delta_{\rm FS}$ is the fine-splitting of the $^2$P$_{3/2}$ and $^2$P$_{1/2}$ levels.}
\label{scheme1} 
\end{figure}

Now we show our proposal for the control of the singlet pairing process. We  assume that in our system there is not only the
weak real magnetic field $B{\bf e}_z$ but also  a laser beam with $\sigma_{+}$- or $\sigma_{-}$-polarization,
which is far off resonant for the D1 and D2 transitions of atom of
both $^{87}$Rb atom and $^{23}$Na atom. 
As shown in Fig.~\ref{scheme1}, this beam can
induce an AC-Stark energy shift $\delta E_{m}^{(j)}$ for the hyperfine
state of a $j$-atom ($j=$Rb, Na) with magnetic quantum number $m$
($m=0,\pm1$). Here we emphasis that the value of $\delta E_{m}^{(j)}$
 depends on both  the  atomic species $j$ and 
magnetic quantum number $m$. The dependence of $\delta E_{m}^{(j)}$
on $m$ is essentially due to the fine splitting bewteen the D1 and
D2 transitions, i.e., the energy gap $\Delta_{{\rm FS}}$ between
the $^{2}{\rm P}_{1/2}$ and $^{2}{\rm P}_{{3}/{2}}$
states of an alkaline atom  \cite{Goldman_2014}. When the detunings of the laser beam for
the D1 and D2 transitions are much larger than $\Delta_{{\rm FS}}$,
for our system $\delta E_{m}^{(j)}$ ($j=$Rb, Na) can be expressed
as 
\begin{eqnarray}
&&\delta E_{m}^{(j)}=S^{(j)}I-\chi V^{(j)}mI,\label{demj}\\
&& (\chi=\pm 1\ {\rm for}\ {\rm laser \ beam\ with}\ \sigma_{\pm}\ {\rm polarization})\nonumber
\end{eqnarray}
with $I$ being the laser intensity. In the right-hand-side of Eq.
(\ref{demj}) the  $m$-independent  term and the linear term of  $m$ are called as
scalar and vector light shifts, respectively. The coefficients
$S^{(j)}$ and $V^{(j)}$ ($j=$Rb, Na) are determined by the laser frequency and the electronic dipole-transition
matrix element of the $j$-atom. If the laser beam has $\sigma_{-}$-polarization,
the result is quite similar. The derivation of Eq. (\ref{demj}) and
the general introduction for the vector light shift can be found in the review article   \cite{Goldman_2014} and the references therein.

Combining Eq. (\ref{demj}) and Eq. (\ref{emj}), we can obtain the
energy of a $j$-atom ($j=$Rb, Na) with magnetic quantum number
$m$ ($m=0,\pm1$) in the presence of both real magnetic field $B$
and laser-induced vector light shift: 
\begin{eqnarray}
E_{m}^{(j)}&=&S^{(j)}I-p^{(j)}m\left(B+B_{{\rm L}}^{(j)}\right)+q^{(j)}m^{2}B^{2},\nonumber\\
&&\ \ \ \ \ \ \ \ \ \ \ \ \ \ \ \ \ \ \ \ \ \ \ \ \ \ \ \ \ \ \ \
(j={\rm Rb},\ {\rm Na}), \label{emj2}
\end{eqnarray}
where the factor $B_{{\rm L}}^{(j)}$ is defined as
\begin{equation}
B_{{\rm L}}^{(j)}=\chi\frac{V^{(j)}}{p^{(j)}}I,\label{beffj}
\end{equation}
and describes the contribution from the vector light shift. It is clear that the effect of the vector light shift is same as the linear Zeeman shift given by a synthetic magnetic field (SMF) $B_{{\rm L}}^{(j)}{\bf e}_z$ ($j$=Na, Rb).

Eqs. (\ref{emj2}) and (\ref{d7}) yield that in the presence of the the laser beam, the detuning
$\Delta_{\left\langle 7\right\rangle }$ for the singlet pairing process
becomes 
\begin{eqnarray}
\Delta_{\left\langle 7\right\rangle } &=&2\left(p^{({\rm Rb})}-p^{({\rm Na})}\right)B\nonumber\\
&&+2\left(p^{({\rm Rb})}B_{{\rm L}}^{({\rm Rb})}-p^{({\rm Na})}B_{{\rm L}}^{({\rm Na})}\right).\label{d73}
\end{eqnarray}
Therefore, the resonance condition $\Delta_{\left\langle 7\right\rangle }=0$
for the singlet pairing process
can be satisfied under a finite real magnetic field, i.e., when
\begin{eqnarray}
B=B_{0}\equiv&-\frac{1}{2}\left(B_{{\rm L}}^{({\rm Rb})}+B_{{\rm L}}^{({\rm Na})}\right)+R\left(B_{{\rm L}}^{({\rm Na})}-B_{{\rm L}}^{({\rm Rb})}\right),\nonumber\\
\label{b0}
\end{eqnarray}
with the ratio $R$ being defined as 
\begin{equation}
R=\frac{p^{({\rm Rb})}+p^{({\rm Na})}}{2\left(p^{({\rm Rb})}-p^{({\rm Na})}\right)}.\label{r}
\end{equation}
As shown above, the values of $p^{({\rm Rb})}$ and $p^{({\rm Na})}$
are very close to each other. Due to this fact, the ratio $R$ in Eq. (\ref{b0})
is very large:
\begin{eqnarray}
R\approx2.03\times10^{3}.\label{rv}
\end{eqnarray}

In the following discussions, for simplicity, we assume that $B_{{\rm L}}^{({\rm Na})}$ is negligibly  small, while  $B_{{\rm L}}^{({\rm Rb})}$ is much larger than $B_{{\rm L}}^{({\rm Na})}$.
This is easy to be realized because $^{87}$Rb and $^{23}$Na atoms have
very different electronic structures \cite{PRL2015Dajun}.
In this case we can only take into account $B_{{\rm L}}^{({\rm Rb})}$ in our calculations.
Thus, according to Eqs. (\ref{b0}) and (\ref{rv}), the resonance point $B_{0}$ for the singlet pairing process is 
\begin{equation}
B_{0}\approx-\left(\frac{1}{2}+R\right)B_{{\rm L}}^{({\rm Rb})}\approx-2030\times B_{{\rm L}}^{({\rm Rb})}.\label{b02}
\end{equation}

\begin{figure}
\includegraphics[width=7cm]{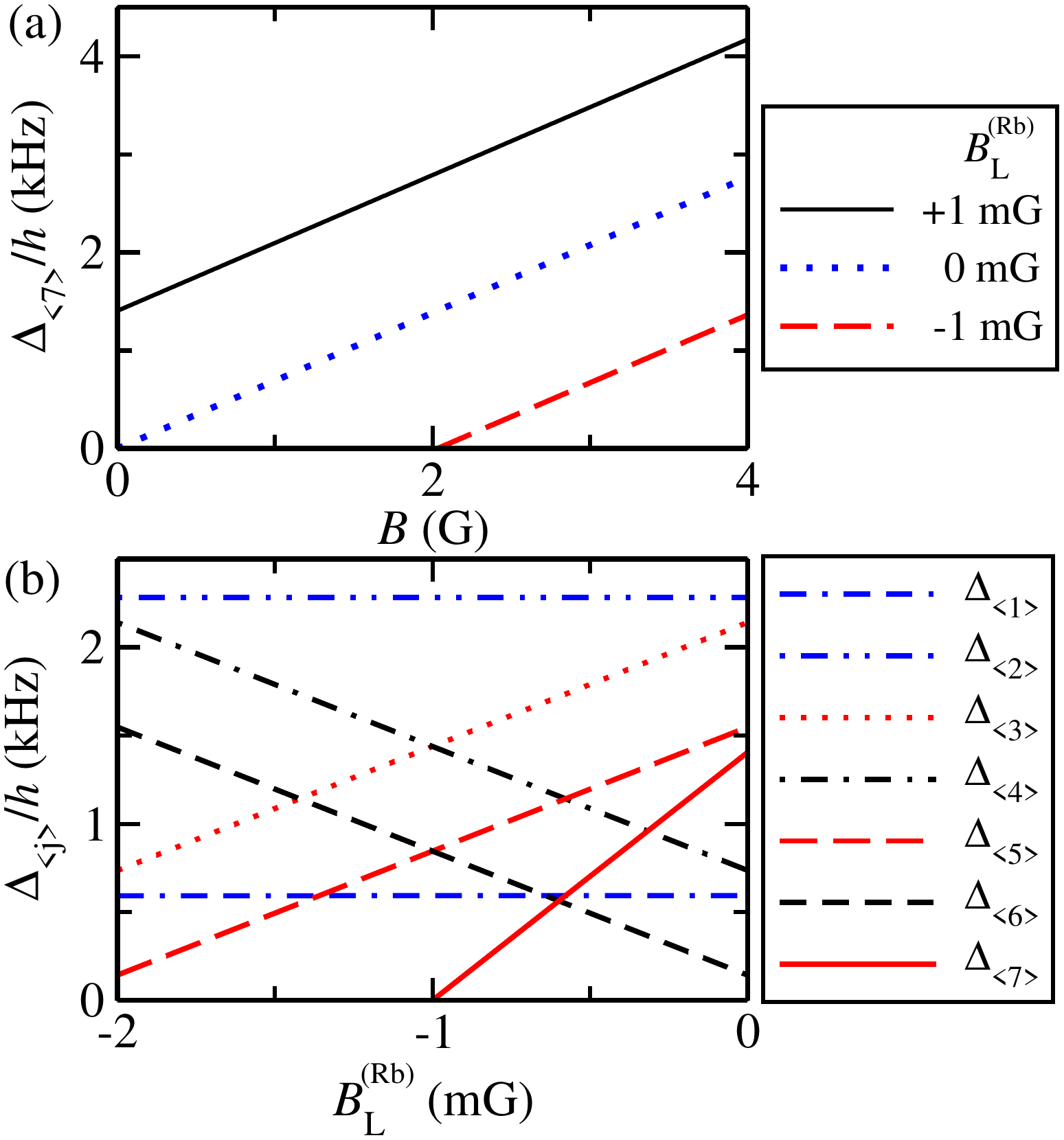}
\caption{(Color online) \textbf{(a):} The detuning $\Delta_{\langle7\rangle}$
as a function of real magnetic field $B$, for the cases with laser-induced SMF {\color{black} $B_{{\rm L}}^{({\rm Na})}=0$ and $B_{{\rm L}}^{({\rm Rb})}=+1$~mG (black solid line), $B_{{\rm L}}^{({\rm Rb})}=0$~mG
(blue dotted line), $B_{{\rm L}}^{({\rm Rb})}=-1$~mG (red dashed
line).} \textbf{(b):} The detunings $\Delta_{\langle1,...7\rangle}$
for all the seven spin-changing processes as a function of $B_{{\rm L}}^{({\rm Rb})}$,  for the cases with $B_{{\rm L}}^{({\rm Na})}=0$ and the real magnetic
field $B$=2.03~G.}
\label{zeeman} 
\end{figure}

In realistic experiments, the laser-induced  SMF $B_{{\rm L}}^{({\rm Rb})}$
is usually of the order of milli Gauss (mG) or even weaker, so that
the laser-induced heating effect is not too strong. Nevertheless,
Eq. (\ref{b02}) shows that  in the presence of this weak SMF
we can realize the resonance condition
$\Delta_{\left\langle 7\right\rangle }=0$ under a  real magnetic field $B_0$ which is as large as several Gauss. On the other hand, at such a large real magnetic
field, the detuning $\Delta_{\langle 1,...,6\rangle}$ for other spin-changing processes can be large enough. Therefore,   the conditions (\ref{cc1}) and (\ref{cc}) can be satisfied simultaneously. This is the basic principle of our proposal. 

We illustrate the above principle in Fig.~\ref{zeeman}.
In Fig.~\ref{zeeman}(a) we show the variation of the detuning
$\Delta_{\langle7\rangle}/h$ with real magnetic field $B$, for the
cases with different laser-induced synthetic magnetic field $B_{{\rm L}}^{({\rm Rb})}$.
It is shown that with the help of $B_{{\rm L}}^{({\rm Rb})}$ we
can realize $\Delta_{\langle7\rangle}=0$ for $B\neq0$. In Fig.~\ref{zeeman}(b) we illustrate the detunings $\Delta_{\langle l\rangle}/h$
($l=1,...,7$) of all spin-changing scattering processes as functions of the
laser-induced magnetic field $B_{{\rm L}}^{({\rm Rb})}$, for the
case with real magnetic field $B$=2.03G. It is clearly shown that
by changing $B_{{\rm L}}^{({\rm Rb})}$ one can tune $\Delta_{\langle7\rangle}$
to be zero while keep the detunings $\Delta_{\langle1,...6\rangle}/h$
of other processes to be as large as several kHz.

\section{Spin oscillation of two trapped atoms}\label{secIV}

In the above section we show our approach for the control of singlet-pairing
process via a light-induced SMF. 
Now we apply this approach on 
a simple system with one
 $^{87}$Rb  and one $^{23}$Na atom. 
We assume these two atoms are confined in an isotropic harmonic trap, e.g., an optical tweezer or a site of 
an optical lattice.
For simplicity, in this section we further assume the confinement has
 the same angular frequency $\omega$ for each atom{\color{black}\cite{MWL,PRA2006MWL}}. 
Thus, the center-of-mass
motion is decoupled with the relative motion, and in our calculation
we can only consider the quantum state of two-body relative motion
and spin. The Hilbert space ${\cal H}$ of our system is given
by ${\cal H}={\cal H}_{r}\otimes{\cal H}_{s{\rm Rb}}\otimes{\cal H}_{s{\rm Na}}$,
with ${\cal H}_{r}$ being the Hilbert space for the two-atom spatial
relative motion, and ${\cal H}_{sj}$ ($j=$Rb, Na) being the one
for the internal state of the $j$-atom. Here we use the symbol $|\rangle\!\rangle$
to denote the state in ${\cal H}$, $|\rangle_{r}$ for state in ${\cal H}_{r}$,
and $|m\rangle_{j}$ ($j=$Rb, Na) for the state in ${\cal H}_{sj}$
with magnetic quantum number $m$.

As in Sec. \ref{secIII}, in our calculation we only take into account the real
magnetic field $B$ and the laser-induced SMF $B_{{\rm L}}^{({\rm Rb})}$ for the $^{87}$Rb atom, and assume the  laser-induced SMF for the $^{23}$Na is negligible.
Accordingly, the
Hamiltonian for our problem is 
\begin{equation}
\hat{H}=\hat{H}_{{\rm ho}}+\hat{U}_{{\rm Rb-Na}}({\bf r})+\hat{Z}\left(B,B_{{\rm L}}^{({\rm Rb})}\right),\label{hh}
\end{equation}
with 
\begin{eqnarray}
\hat{H}_{{\rm ho}}=\frac{\hat{{\bf p}}^{2}}{2\mu}+\frac{\mu}{2}\omega^2 {\bf r}^{2},\label{hho}
\end{eqnarray}
where $\hat{{\bf p}}$ is the relative-momentum operator of the two atoms,
$\mu$ and ${\bf r}$ are the reduced mass and relative
position of these two atoms as before, and the inter-species interaction
$\hat{U}_{{\rm Rb-Na}}({\bf r})$ is given by Eq. (\ref{uab}). Here the
${\bf r}$-independent operator $\hat{Z}(B,B_{{\rm L}}^{({\rm Rb})})$
describes the influence of the real and synthetic magnetic field on
the energy of atomic spin states, and can be expressed as 
\begin{eqnarray}
 &  & \hat{Z}\left(B,B_{{\rm L}}^{({\rm Rb})}\right)\nonumber \\
 & = & \sum_{m,m'}\left(E_{m}^{({\rm Rb})}+E_{m'}^{({\rm Na})}\right)|m\rangle_{{\rm Rb}}\langle m|\otimes|m'\rangle_{{\rm Na}}\langle m'|,\nonumber \\
\label{zz}
\end{eqnarray}
where both $m$ and $m'$ can take value $(0,\pm1)$ and the energy
$E_{m}^{(j)}$ ($j=$Rb, Na) is given by Eq. (\ref{emj2}).

\begin{figure}[t]
\includegraphics[width=8.0cm]{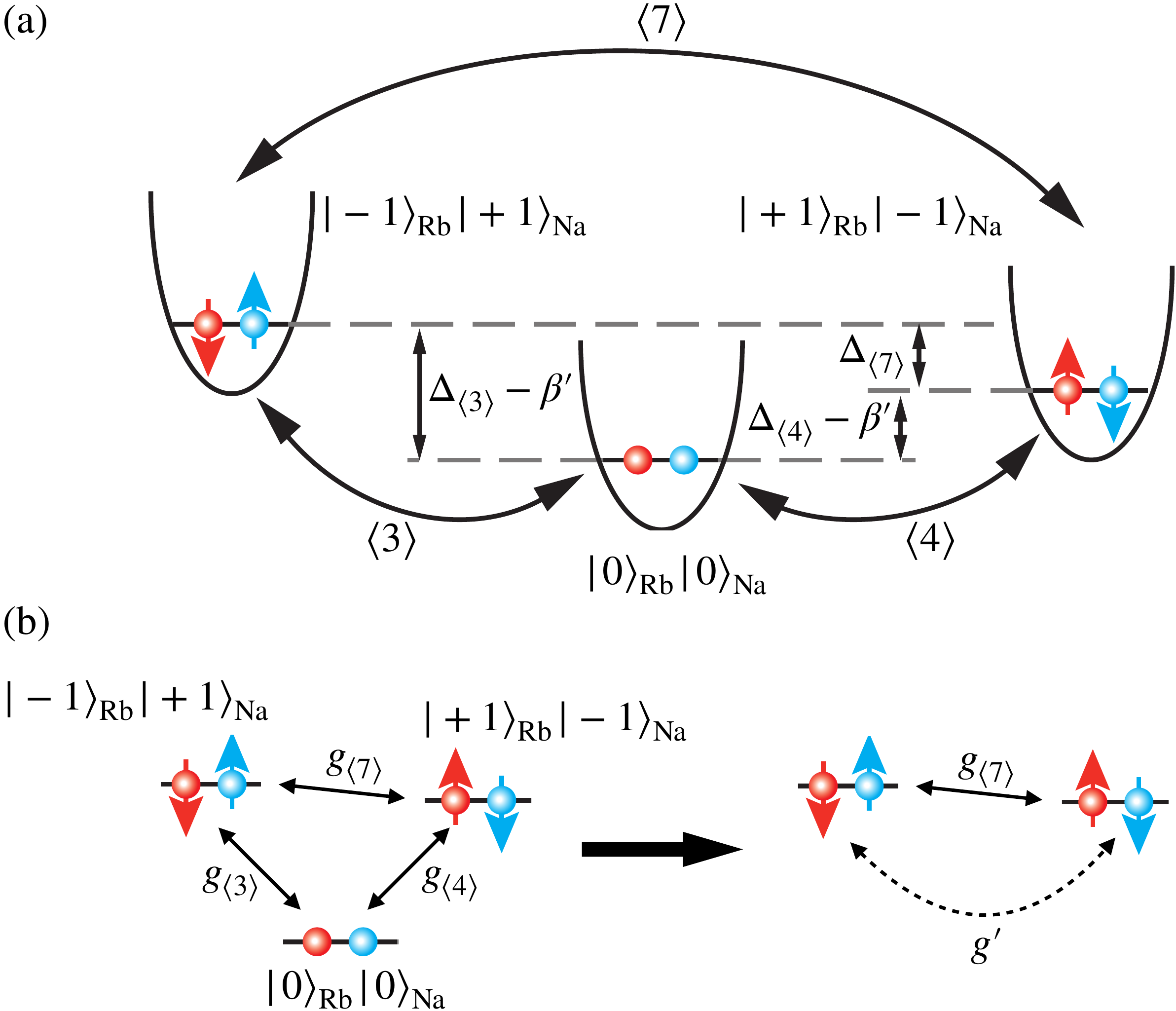} \caption{(Color online) {\bf (a):} A schematic illustration for the spin dynamics of a
$^{87}$Rb atom (red ball) and a $^{23}$Na atom (blue ball) confined
in an isotropic Harmonic trap. Here we also show the effective detunings $\Delta_{\langle 3,4\rangle}-\beta^\prime$ for the processes $\langle 3,4\rangle$ and $\Delta_{\langle 7\rangle}$ for the process $\langle 7\rangle$ (see Sec.~\ref{secIV}). {\bf (b):} 
Under the condition (\ref{cc2}) the state $|0\rangle_{\rm Rb}|0\rangle_{\rm Na}$ can be adiabatically eliminated. In this case the states
$ |-1\rangle_{\rm Rb}|+1\rangle_{\rm Na}$ and $ |+1\rangle_{\rm Rb}|-1\rangle_{\rm Na}$
experiences both the direct coupling $g_{\langle 7\rangle}$ and an effective coupling $g^\prime$ which is given by the virtual transitions  between  these two states and $ |0\rangle_{\rm Rb}|0\rangle_{\rm Na}$ (i.e., the off-resonant processes $\langle 3,4\rangle$). }
\label{trap} 
\end{figure}

Furthermore, we assume the relative wave
function of the two atoms is initially prepared in the ground state
$|0\rangle_{r}$ of the Hamiltonian $\hat{H}_{{\rm ho}}$ defined in Eq. (\ref{hho}). The interaction $\hat{U}_{{\rm Rb-Na}}({\bf r})$ can induce the transition
between $|0\rangle_{r}$ and the excited states of $\hat{H}_{{\rm ho}}$.
Nevertheless, the direct calculation shows that these transitions
can be neglected when the inter-atomic scattering length $a_{{\rm Rb-Na}}^{(0,1,2)}$
is much smaller than the characteristic length 
\begin{eqnarray}
a_{{\rm ho}}\equiv\sqrt{\frac{\hbar}{\mu\omega}}
\end{eqnarray}
of the trap \cite{trap1}. As shown above, the values
of $a_{{\rm Rb-Na}}^{(0,1,2)}$ are less than $100a_{0}$, while in
almost all of the realistic experiments the confinement characteristic
length $a_{{\rm ho}}$ is larger than $1000a_{0}$. Therefore, in
the lowest-order calculation we can neglect the transition between
different spatial states, i.e., assume  the relative motion of the atoms is frozen in the state $|0\rangle_r$.


In addition, the inter-atomic interaction $\hat{U}_{{\rm Rb-Na}}({\bf r})$
can also induce the transtion between different hypferfine-spin states
via the spin-changing processes shown in Sec. \ref{secII}. Here we consider the
system where the $^{87}$Rb atom and $^{23}$Na atom are prepared
in hyperfine spin states $|-1\rangle_{{\rm Rb}}$ and $|+1\rangle_{{\rm Na}}$,
respectively. In this case, only the processes $\langle3\rangle$,
$\langle4\rangle$ and $\langle7\rangle$ can occur during the evolution,
as shown in Fig.~\ref{trap}(a). As a result, the state at time $t$ can be expressed as 
\begin{equation}
|\Psi(t)\rangle\!\rangle=\sum_{\left\{ m,m'\right\} }C_{m,m'}(t)|0\rangle_{r}\otimes|m\rangle_{{\rm Rb}}\otimes|m'\rangle_{{\rm Na}},\label{psit}
\end{equation}
where $\left\{ m,m'\right\} $ can take the values $\left\{ -1,+1\right\} $, $\left\{0,0\right\}$ and $\left\{ +1,-1\right\} $. 
Furthermore, by projecting the Schr$\ddot{{\rm o}}$dinger
equation in the subspace spanned by the three states $|0\rangle_{r}\otimes|m\rangle_{{\rm Rb}}\otimes|m'\rangle_{{\rm Na}}$ involved in Eq. (\ref{psit}), we can obtain the equation for the coefficients $C_{m,m'}(t)$ (up to a global phase factor):
\begin{equation}
i\hbar\frac{d}{dt}\left(\begin{array}{c}
C_{-1,+1}\\
C_{0,0}\\
C_{+1,-1}
\end{array}\right)={\cal M}\left(\begin{array}{c}
C_{-1,+1}\\
C_{0,0}\\
C_{+1,-1}
\end{array}\right),\label{equa}
\end{equation}
where the matrix ${\cal M}$ is given by 
\begin{equation}
{\cal M}=\left(\begin{array}{ccc}
-\beta' & g_{\langle3\rangle} & g_{\langle7\rangle}\\
g_{\langle3\rangle} &-\Delta_{\langle3\rangle} & g_{\langle4\rangle}\\
g_{\langle7\rangle} & g_{\langle4\rangle} & -\beta'-\Delta_{\langle7\rangle}
\end{array}\right),\label{calm}
\end{equation}
with the detunings $\Delta_{\langle3,7\rangle}$
being defined in Sec. \ref{secIII}.
In Eq. (\ref{calm}) the {\color{black} characteristic interaction strength (effective spin-changing intensity)} $g_{\langle3,4,7\rangle}$ are given by the projection of the inter-atomic interaction $\hat{U}_{{\rm Rb-Na}}({\bf r})$ on the ground state $|0\rangle_r$ of the relative spatial motion, and can be expressed as
\begin{eqnarray}
g_{\langle3\rangle} & = & g_{\langle4\rangle}=\beta'-\gamma'/3;\label{ge1}\\
g_{\langle7\rangle} & = & \gamma'/3.\label{ge2}
\end{eqnarray}
with
\begin{eqnarray}
 (\alpha', \beta',\gamma')=\frac{1}{\pi^{3/2}a_{{\rm ho}}^{3}}(\alpha, \beta,\gamma).\label{ge3}
\end{eqnarray}

With the help of the above equations we can obtain a clear qualitative
understanding for the spin dynamics of these two atoms. In our system each of the spin-changing  process $\langle 3,4,7\rangle$ can induce a quantum transition between two spin states (Fig.~\ref{trap}), i.e.,
\begin{eqnarray}
\langle 3\rangle \ {\rm induces}:\ \ \  |-1\rangle_{\rm Rb}|+1\rangle_{\rm Na}&\leftrightarrow&
|0\rangle_{\rm Rb}|0\rangle_{\rm Na};\nonumber\\
\langle 4\rangle\ {\rm induces}:\ \ \  |+1\rangle_{\rm Rb}|-1\rangle_{\rm Na}&\leftrightarrow&
|0\rangle_{\rm Rb}|0\rangle_{\rm Na};\nonumber\\
\langle 7\rangle\ {\rm induces}:\ \ \  |+1\rangle_{\rm Rb}|-1\rangle_{\rm Na}&\leftrightarrow&
|-1\rangle_{\rm Rb}|+1\rangle_{\rm Na}.\nonumber
\end{eqnarray}
Furthermore, using Eq. (\ref{calm}) and the fact $\Delta_{\langle4\rangle}=\Delta_{\langle3\rangle}-\Delta_{\langle7\rangle}$, we find that the  detuning $\delta_{\langle l \rangle}$  of the   transition induced by the process $\langle l\rangle$ ($l=3,4,7$) is given by 
\begin{eqnarray}
\delta_{\langle 3,4\rangle}=\Delta_{\langle 3,4\rangle}-\beta^\prime
\end{eqnarray}
and
\begin{eqnarray}
\delta_{\langle 7\rangle}=\Delta_{\langle 7\rangle},
\end{eqnarray}
while the direct coupling intensity  corresponding to this transition is just $g_{\langle l\rangle}$. 
Thus, this transition is significant when $|\delta_{\langle l\rangle}|\ll|g_{\langle l\rangle}|$, and becomes negligible when $|\delta_{\langle l\rangle}|\gg|g_{\langle l\rangle}|$. As
shown in Sec. \ref{secIII}, this can be realized with the help of
the light-induced SMF $B_{{\rm L}}^{({\rm Rb})}$ via our approach.

\begin{figure*}[t]
\includegraphics[width=12cm]{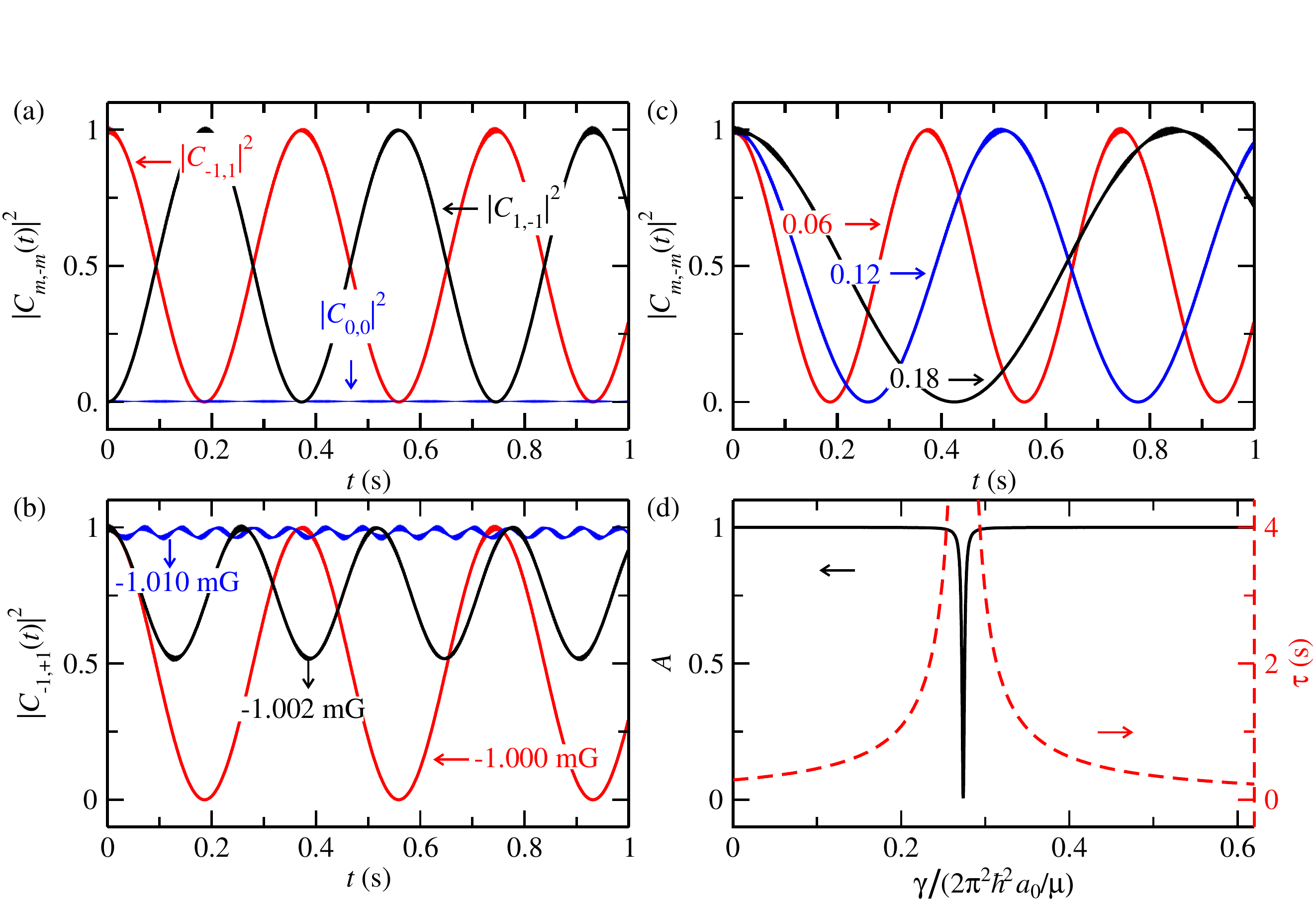}   
\caption{(Color online) 
{\bf (a):}  $|C_{m,m^\prime}(t)|^2$ ($m,m^\prime=0,\pm 1$) given by the solution of Eq. (\ref{equa}) for initial spin state $|-1\rangle_{\rm Rb}|+1\rangle_{\rm Na}$. Here we consider the system with $\omega=2\pi\times$40 kHz, $(\alpha,\beta,\gamma)=\frac{2\pi\hbar^{2}a_{0}}{\mu}(78.9,-2.5,0.06)$, $B=2.03$G, $B_{\rm L}^{\rm (Rb)}=-1$~mG and $B_{\rm L}^{\rm (Na)}=0$. 
{{\bf (b):}  $|C_{-1,+1}(t)|^2$ for the system of (a), with $B_{\rm L}^{\rm (Rb)}=-1$~mG (red line), $-1.002$~mG (black line),  $-1.010$~mG (blue line).} {\bf (c):}  $|C_{-1,+1}(t)|^2$ for the system of (a), with {$\gamma=(0.06)\frac{2\pi\hbar^{2}a_{0}}{\mu}$ (red line), $(0.12)\frac{2\pi\hbar^{2}a_{0}}{\mu}$ (blue line), $(0.18)\frac{2\pi\hbar^{2}a_{0}}{\mu}$ (black line). 
{\bf (d):} The amplitude $A$ and period $\tau$ of the time oscillation of the profile of $|C_{-1,+1}(t)|^2$ as a function of the interaction parameter $\gamma$. Other parameters are same as (a).}
}
\label{twobody1}
\end{figure*}

According to the above discussion, to enhance the singlet-pairing process $\langle 7\rangle$ and simultaneously suppress the direct effect of
the other two processes, one requires to make 
\begin{eqnarray}
|\delta_{\langle7\rangle}|\ll|g_{\langle7\rangle}|;\ \ \ 
|\delta_{\langle3,4\rangle}|\gg|g_{\langle3,4\rangle}|.\label{cc2}
\end{eqnarray}
Explicitly, under this condition the state $|0\rangle_{\rm Rb}|0\rangle_{\rm Na}$ can be adiabatically eliminated and the
 coefficients $C_{-1,+1}(t)$ and $C_{1,-1}(t)$ satisfy the effective Schr${\ddot {\rm o}}$dinger equation  (up to a constant)
\begin{eqnarray}
i\hbar\frac{d}{dt}\left(\begin{array}{c}
C_{-1,+1}\\
C_{+1,-1}
\end{array}\right)=
\left(
\begin{array}{cc}
\delta_{\langle 7\rangle} &  g_{\langle7\rangle}+g^\prime\\
g_{\langle7\rangle}+g^\prime & 0
\end{array}
\right)
\left(
\begin{array}{cc}
C_{-1,+1}\\
C_{+1,-1}
\end{array}\right),\nonumber\\
\label{ese}
\end{eqnarray}
with 
\begin{eqnarray}
g^\prime=-\frac{g_{\langle3\rangle}^2}{\delta_{\langle3\rangle}}.
\end{eqnarray}
Here we have used the facts $g_{\langle 3\rangle}=g_{\langle 4\rangle}$ and $\delta_{\langle 3\rangle}\approx\delta_{\langle 4\rangle}$. Eq. (\ref{ese}) shows that in addition to the direct coupling  $g_{\langle7\rangle}$ induced by the process $\langle 7\rangle$, the states
$ |-1\rangle_{\rm Rb}|+1\rangle_{\rm Na}$ and $ |+1\rangle_{\rm Rb}|-1\rangle_{\rm Na}$
also experiences an effective coupling $g^\prime$. This term is given by the virtual transitions between  these two states and the far-off resonant state $ |0\rangle_{\rm Rb}|0\rangle_{\rm Na}$, and is actually an indirect effect of the processes 
$\langle 3,4\rangle$ (Fig.~\ref{trap}(b)). When $|g_{\langle7\rangle}|$ is comparable with or much larger than $|g^\prime|$, the spin-changing process $\langle 7\rangle$ makes considerable contribution for the 
the Rabi oscillation between $ |-1\rangle_{\rm Rb}|+1\rangle_{\rm Na}$ and $ |+1\rangle_{\rm Rb}|-1\rangle_{\rm Na}$. Therefore, one can observe the effect of the process $\langle 7\rangle$ and measure the  corresponding interaction parameter $\gamma$ with the help of this Rabi oscillation.

As an example, we consider the case with trapping frequency $\omega=2\pi\times 40$~kHz and a real magnetic field $B=2.03$G. According to the calculation in the above section, for such a system when the laser-induced SMF is tuned to  $B_{\rm L}^{\rm (Rb)}=-1$mG, we have $\delta_{\langle 7\rangle}/h=0.022{\rm Hz}=0.054g_{\langle 7\rangle}/h$ and $|\delta_{\langle3,4\rangle}|=29|g_{\langle3,4\rangle}|$. Therefore, in this case the condition (\ref{cc2}) is satisfied and thus the Rabi oscillation 
$ |-1\rangle_{\rm Rb}|+1\rangle_{\rm Na}\leftrightarrow |+1\rangle_{\rm Rb}|-1\rangle_{\rm Na}$ can be enhanced. To illustrate this effect, we exactly solve the three-level Schr${\rm{\ddot o}}$dinger equation (\ref{equa}) for the initial spin state $|-1\rangle_{\rm Rb}|+1\rangle_{\rm Na}$, and show the time-evolution of the populations  for each spin state in Fig.~\ref{twobody1}(a).
It is clearly shown that  the amplitudes of  
this Rabi oscillation (i.e., the amplitudes of the time oscillations of $|C_{-1,+1}(t)|^2$ and $|C_{+1,-1}(t)|^2$) is almost unit, while the population of the state $|0\rangle_{\rm Rb}|0\rangle_{\rm Na}$ is almost zero. In Fig.~\ref{twobody1}(b), we show the time oscillation of the population $|C_{-1,1}(t)|^2$ for various other values of the SMF $B_{\rm L}^{\rm (Rb)}$. It is clearly shown that both the period  and amplitude of the oscillations are extremely sensitive to  $B_{\rm L}^{\rm (Rb)}$.

Our further calculation shows that in this system we have 
$g^\prime=-4.3g_{\langle 7\rangle}$. Therefore, 
the singlet-pairing process $\langle 7\rangle$ makes considerable and observable contribution for the  above Rabi oscillation, although it is still less than the contribution from
 the indirect coupling $g^\prime$ because of the extremely-weak interaction parameter $\gamma$ of the $^{23}{\rm Na}$-$^{87}{\rm Rb}$ mixture ($\gamma\approx0.024\beta=(0.06)\frac{2\pi\hbar^{2}a_{0}}{\mu}$, as shown in Sec.~II). in Fig.~\ref{twobody1}(c), we show the time oscillation of the population $|C_{-1,1}(t)|^2$ for various other values of $\gamma$, which are near the above realistic one. It is clearly shown that the period $\tau$ of this oscillation seriously depends on $\gamma$. Therefore, one can precisely measure the value of  $\gamma$ by detecting $\tau$. 
In Fig.~\ref{twobody1}(d), we further illustrate the period $\tau$ and the amplitude $A$ of the Rabi oscillation $ |-1\rangle_{\rm Rb}|+1\rangle_{\rm Na}\leftrightarrow |+1\rangle_{\rm Rb}|-1\rangle_{\rm Na}$  as functions of $\gamma$, respectively. It is shown that 
if  $\gamma$ were taking a particular value $\gamma_\ast\equiv(0.26)\frac{2\pi\hbar^{2}a_{0}}{\mu}$, 
we would have $\tau=\infty$ and $A=0$,  i.e., this Rabi oscillation would be totally suppressed. That is just because for our system the coupling parameters $g_{\langle 7\rangle}$ and $g^\prime$ satisfy $g_{\langle 7\rangle}+g^\prime=0$ for $\gamma=\gamma_\ast$. Namely, there is a completely destructive interference between the direct and indirect transitions from $ |+1\rangle_{\rm Rb}|-1\rangle_{\rm Na}$ to $ |-1\rangle_{\rm Rb}|+1\rangle_{\rm Na}$, which are induced by the singlet-pairing process $\langle 7\rangle$ and the virtual processes through  $|0\rangle_{\rm Rb}|0\rangle_{\rm Na}$, respectively. In the region around $\gamma=\gamma_\ast$, the period $\tau$ and the amplitude $A$ sensitively dependent on the value of $\gamma$, and thus can be used for the precise measurement this interaction parameter \cite{trap2}. 

{\color{black}In the end of this section, we emphasis that the systems of two ultracold atoms confined in a single trap have been prepared in many experiments with optical lattice site or optical tweezer  \cite{Guan2019OT,Anderegg2019OT,Liu2018OT}, and has been used for the inter-atomic interaction intensity for alkaline-earth (like) atoms  \cite{Hofer2015IM,Cappellini2019trap}. Therefore, the spin-oscillation processes studied in this section are also possibly to be realized in current experiments, with which one can observe  the singlet-pairing and measure the corresponding interaction intensity $\gamma$, as shown above. In addition, using these processes one can also prepare the two-atom entangled states, {\it e.g.}, the Bell state $\left(|-1\rangle_{\rm Na}|+1\rangle_{\rm Rb}\pm i|+1\rangle_{\rm Na}|-1\rangle_{\rm Rb}\right)/\sqrt{2}$ which can be used for for the studies of quantum information or quantum physics. }

\section{Binary mixtures of spin-1 BECs}\label{secV}

\begin{figure*}[t]
\centering \includegraphics[width=11cm]{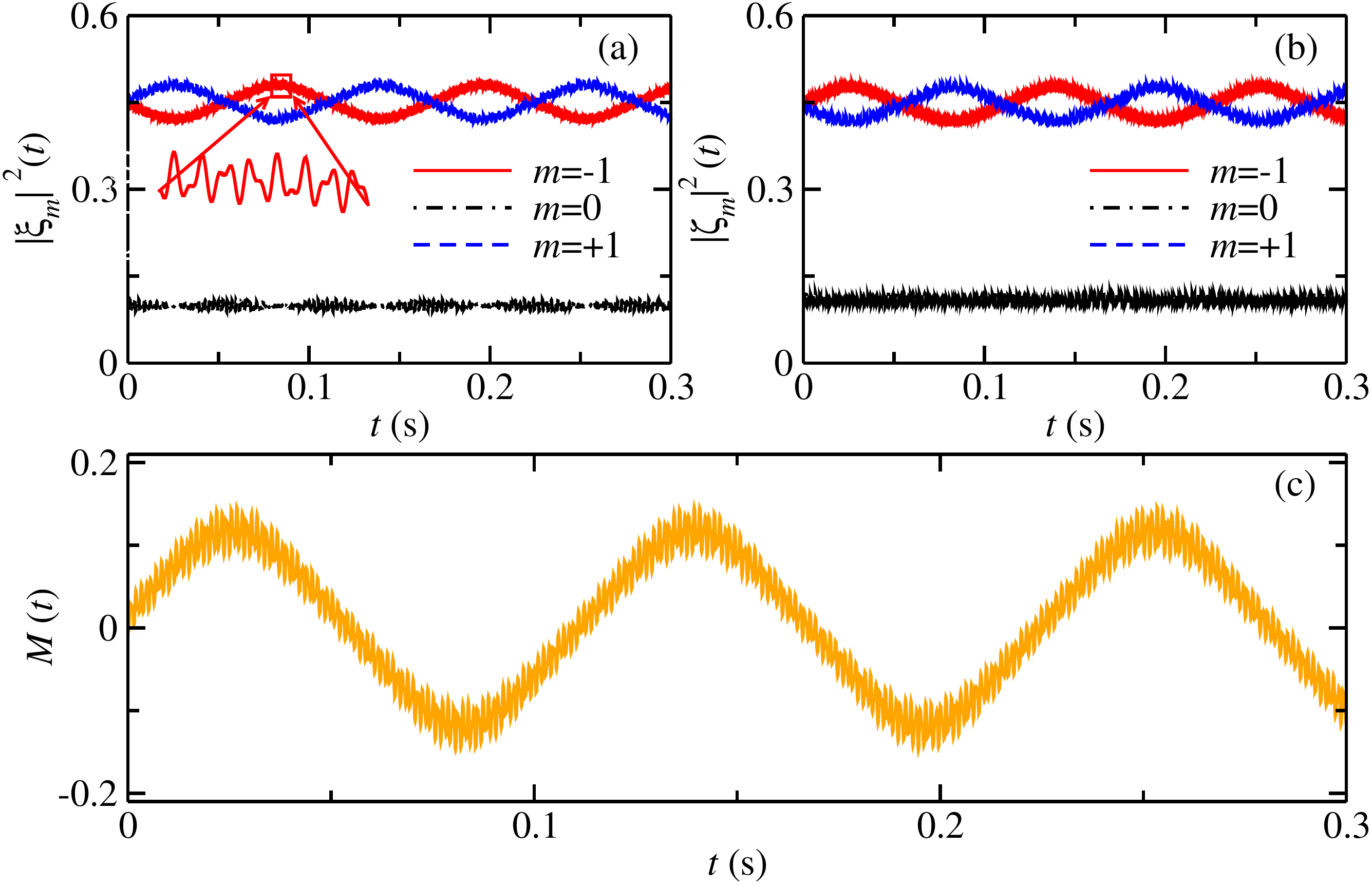} \caption{(Color online) {\bf (a):} The time evolution of the populations $|\xi_m(t)|^2$ and  $|\zeta_m(t)|^2$ ($m=0,\pm 1$), which are given by the numerical solution of Eqs. (\ref{cp1}-\ref{cp4}) under the initial conditions (\ref{i1}, \ref{i2}). Other details of our calculation are shown in the main text of Sec. V. {\bf (b):} The time evolution of the relative magnetization $M(t)$ for the system of (a).}
\label{fig6}
\end{figure*}

Now we study the control of the singlet-pairing process in  a two-species  BEC of spin-1  $^{87}$Rb and $^{23}$Na atoms with our approach. As in the above sections, we assume our system is confined in an isotropic harmonic trap, and  there are both a real magnetic field and a laser-induced SMF. As shown in Appendix \ref{appendix1}, under the mean-field and single-mode approximations  \cite{PRL1998Bigelow,PRA1999Pu,PRA2002_SMA:Yi,NJPZhang_2003,PR2012_Spinor_BEC:Ueda,RMP2013_Spinor_BEC:Ueda,Jie1,Jie2}, the states of $^{87}$Rb and $^{23}$Na BECs can be described by  three-component wave functions 
\begin{eqnarray}
\psi(\textbf{r})
\left(
\begin{array}{l}
\xi_{-1}(t),\\
\xi_{0}(t),\\
\xi_{+1}(t)
\end{array}
\right)
\ \ \ 
{\rm and}
\ \ \ 
\phi(\textbf{r})
\left(
\begin{array}{l}
\zeta_{-1}(t),\\
\zeta_{0}(t),\\
\zeta_{+1}(t)
\end{array}
\right),\label{wff}
\end{eqnarray}
respectively. The spatial wave functions $\psi(\textbf{r})$ and $\phi(\textbf{r})$, which are normalized to unit, are determined by the Gross-Pitaevskii equations (\ref{cgpmu1}, \ref{cgpmu}).
 In addition, the spin states of the $^{87}$Rb and $^{23}$Na BECs are described by the time-dependent  complex row vectors $[\xi_{-1}(t),\xi_{0}(t),\xi_{+1}(t)]^{\rm T}$ and $[\zeta_{-1}(t),\zeta_{0}(t),\zeta_{+1}(t)]^{\rm T}$, respectively, which satisfy the normalization condition 
\begin{eqnarray}
\sum_{m=0,\pm 1}|\xi_{m}|^{2}=\sum_{m=0,\pm 1}|\zeta_{m}|^{2}=1.
\end{eqnarray}
The time evolution of the components $\xi_m(t)$ and $\zeta_m(t)$ ($m=0,\pm 1$), i.e., the spin dynamics of the $^{87}$Rb and $^{23}$Na BECs, are determined by the equations (Appendix  \ref{appendix1}):
\begin{widetext}
\begin{eqnarray}
\imath\hbar\dot{\xi}_{\pm1} & = & E_{\pm1}^{(\rm Rb)}\xi_{\pm1}+\bar{\beta}_{\rm Rb}\left[\left(|\xi_{\pm1}|^{2}+|\xi_{0}|^{2}-|\xi_{\mp1}|^{2}\right)\xi_{\pm1}+\xi_{0}^{2}\xi_{\mp1}^{\ast}\right]\nonumber \\
 &  &+ \bar{\beta}^{(\rm Na)}\left[\left(|\zeta_{\pm1}|^{2}-|\zeta_{\mp1}|^{2}\right)\xi_{\pm1}+\zeta_{\pm1}\xi_{0}\zeta_{0}^{\ast}+\zeta_{0}\xi_{0}\zeta_{\mp1}^{\ast}\right]+ \frac{\bar{\gamma}^{(\rm Na)}}{3}\left(\zeta_{\mp1}\xi_{\pm1}+\zeta_{\pm1}\xi_{\mp1}-\zeta_{0}\xi_{0}\right)\zeta_{\mp1}^{\ast};\label{cp1}\\
\imath\hbar\dot{\xi}_{0} & = &E_{0}^{(\rm Rb)}\xi_{0}+ \bar{\beta}_{\rm Rb}\left[\left(|\xi_{+1}|^{2}+|\xi_{-1}|^{2}\right)\xi_{0}+2\xi_{+1}\xi_{-1}\xi_{0}^{\ast}\right]\nonumber \\
 &  &+ \bar{\beta}^{(\rm Na)}\left[\zeta_{0}\xi_{+1}\zeta_{+1}^{\ast}+\zeta_{0}\xi_{-1}\zeta_{-1}^{\ast}+\left(\zeta_{-1}\xi_{+1}+\xi_{-1}\zeta_{+1}\right)\zeta_{0}^{\ast}\right]+\frac{\bar{\gamma}^{\rm (Na)}}{3}\left(-\zeta_{-1}\xi_{+1}+\zeta_{0}\xi_{0}-\zeta_{+1}\xi_{-1}\right)\zeta_{0}^{\ast};\\
\imath\hbar\dot{\zeta}_{\pm1} & = & E_{\pm1}^{(\rm Na)}\zeta_{\pm1}+\bar{\beta}_{\rm Na}\left[\left(|\zeta_{\pm1}|^{2}+|\zeta_{0}|^{2}-|\zeta_{\mp1}|^{2}\right)\zeta_{\pm1}+\zeta_{0}^{2}\zeta_{\mp1}^{\ast}\right]\nonumber \\
 &  &+ \bar{\beta}^{(\rm Rb)}\left[\left(|\xi_{\pm1}|^{2}-|\xi_{\mp1}|^{2}\right)\zeta_{\pm1}+\xi_{\pm1}\zeta_{0}\xi_{0}^{\ast}+\xi_{0}\zeta_{0}\xi_{\mp1}^{\ast}\right]+ \frac{\bar{\gamma}^{(\rm Rb)}}{3}\left(\zeta_{\mp1}\xi_{\pm1}+\xi_{\mp1}\zeta_{\pm1}-\zeta_{0}\xi_{0}\right)\xi_{\mp1}^{\ast};\\
\imath\hbar\dot{\zeta}_{0} & = & E_{0}^{(\rm Na)}\zeta_{0}+\bar{\beta}_{\rm Na}\left[\left(|\zeta_{+1}|^{2}+|\zeta_{-1}|^{2}\right)\zeta_{0}+2\zeta_{+1}\zeta_{-1}\zeta_{0}^{\ast}\right]\nonumber \\
 & &+ \bar{\beta}^{(\rm Rb)}\left[\xi_{0}\zeta_{+1}\xi_{+1}^{\ast}+\xi_{0}\zeta_{-1}\xi_{-1}^{\ast}+\left(\xi_{-1}\zeta_{+1}+\xi_{+1}\zeta_{-1}\right)\xi_{0}^{\ast}\right]+ \frac{\bar{\gamma}^{(\rm Rb)}}{3}\left(-\zeta_{-1}\xi_{+1}+\xi_{0}\zeta_{0}-\zeta_{+1}\xi_{-1}\right)\xi_{0}^{\ast},\label{cp4}
\end{eqnarray}
\end{widetext}
with the effective interaction strengths being defined as
\begin{eqnarray}
\bar{\beta}_{\rm Rb}&=&\beta_{\rm Rb}N_{\rm Rb}\int d\textbf{r}|\psi(\textbf{r})|^{4},\\
\bar{\beta}_{\rm Na}&=&\beta_{\rm Na}N_{\rm Na}\int d\textbf{r}|\phi(\textbf{r})|^{4},\\
\bar{\beta}^{(\rm Rb, Na)}&=&\beta N_{\rm Rb,Na}\int d\textbf{r}|\psi(\textbf{r})|^{2}|\phi(\textbf{r})|^{2},\\
\bar{\gamma}^{(\rm Rb, Na)}&=&\gamma N_{\rm Rb,Na}\int d\textbf{r}|\psi(\textbf{r})|^{2}|\phi(\textbf{r})|^{2}.
\end{eqnarray}
Here $N_j$ ($j={\rm Na},\ {\rm Rb}$) are the number of the $j$-atoms.
In Eqs. (\ref{cp1}-\ref{cp4}) the effective Zeeman energies $E_{0,\pm 1}^{(j)}$ ($j=$Na, Rb) are functions of  the real magnetic field $B$ and the laser-induced SMF  $B_{\rm L}^{(\rm Na,Rb)}$, as defined in Eq. (\ref{emj2}).
Therefore, one can control these energies, and thus the detunings $\Delta_{\langle l\rangle}$ ($l=1,...,7$) for the spin-changing process, via $B$ and  $B_{\rm L}^{(\rm Na,Rb)}$.
As mentioned in Sec. I, we can enhance the singlet-pairing process $\langle 7\rangle$ by tuning $B$ and  $B_{\rm L}^{(\rm Na,Rb)}$ to the proper values where this process is resonant while the other spin-changing processes are far-off resonant, i.e., the conditions (\ref{cc1}, \ref{cc}) are satisfied.

\begin{figure}[t]
\centering \includegraphics[width=7.cm]{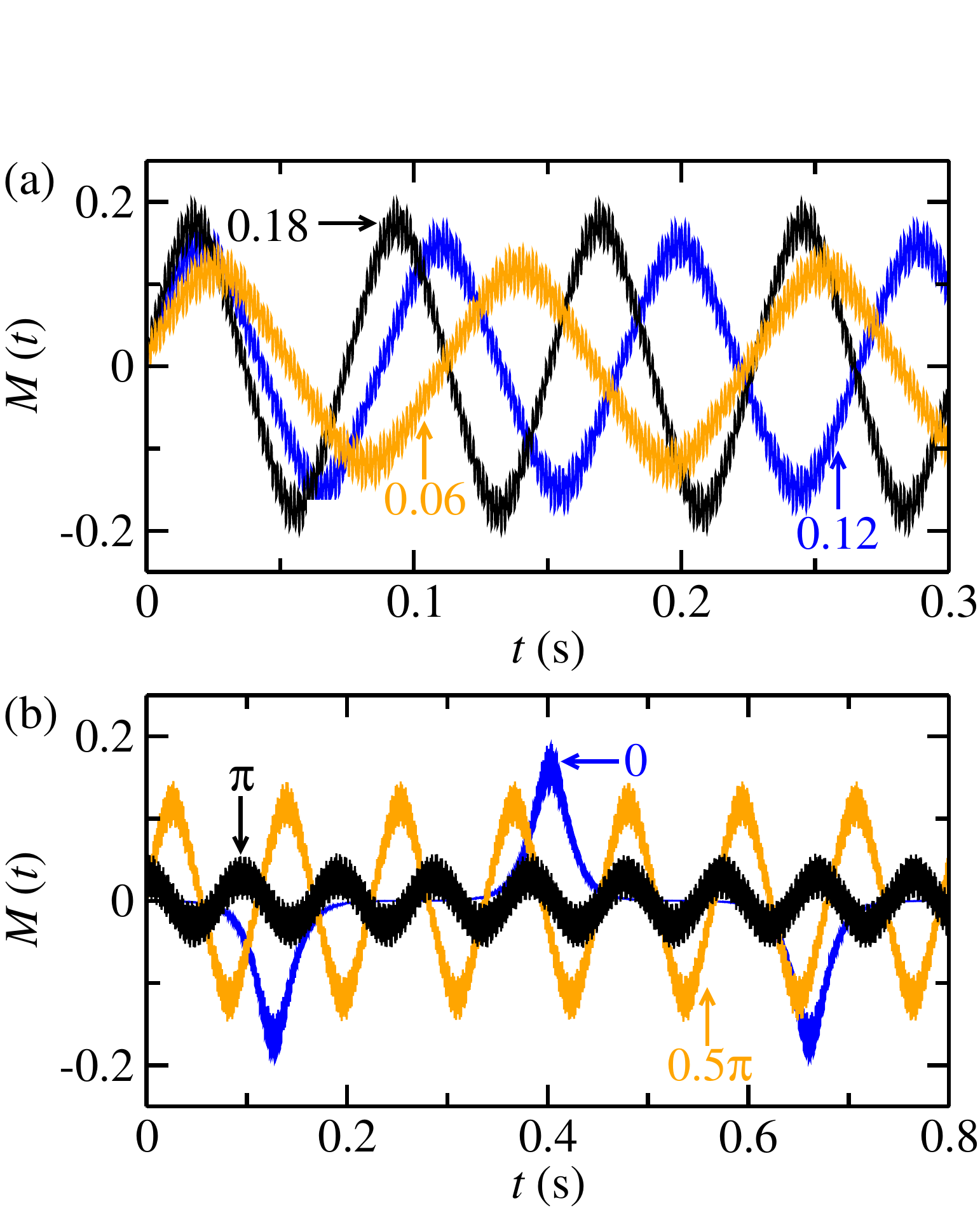} \caption{(Color online) The time evolution of the relative magnetization $M(t)$.  {\bf (a):} Cases with $\gamma=(0.06)\frac{2\pi\hbar^{2}a_{0}}{\mu}$~(orange line), $\gamma=(0.12)\frac{2\pi\hbar^{2}a_{0}}{\mu}\pi$ (blue line),  and $\gamma=(0.18)\frac{2\pi\hbar^{2}a_{0}}{\mu}$ (black line). Other parameters are same as Fig. (\ref{fig6}).
  {\bf (b):} Cases with initial relative phase factor $\arg\left[\zeta_{-1}(t=0)\right]=0$~(blue line), $\arg\left[\zeta_{-1}(t=0)\right]=0.5\pi$ (orange line) and $\arg\left[\zeta_{-1}(t=0)\right]=\pi$ (black line). Other parameters are same as Fig. (\ref{fig6}).}
\label{fig8}
\end{figure}

To illustrate the above technique, we investigate the spin dynamics for the case with $B$=2.03G, $B_{\rm L}^{\rm (Rb)}$=-1mG and $B_{\rm L}^{\rm (Na)}$ is negligible, where the conditions (\ref{cc1}, \ref{cc}) can be satisfied. In our calculation we assume $~\omega_{\rm Rb}=2\pi\times200$ Hz, $\omega_{\rm Na}=2\pi\times500$ Hz, and $N_{\rm Rb}=N_{\rm Na}=4\times 10^4$,
and derive the atomic  probability  densities $|\psi({\bf r})|^2$ and $|\phi({\bf r})|^2$ via the Thomas-Fermi approximation (Appendix B).
 As a result, the effective spin-singlet pairing interaction strength is ${\bar \gamma}^{(\rm Rb,Na)}/h=0.965$ Hz and the corresponding chemical potentials are $\mu_{\rm Rb}/h=3970$ Hz and $\mu_{\rm Na}/h=6093$ Hz (see Eqs. (\ref{cgpmu1}-\ref{cgpmu})). We numerically  solve Eqs. (\ref{cp1})-(\ref{cp4}) for this system, under the initial condition 
\begin{eqnarray}
\left(
\begin{array}{l}
\xi_{-1}(t=0)\\
\xi_{0}(t=0)\\
\xi_{+1}(t=0)
\end{array}
\right)
&=&
\left(
\begin{array}{l}
\sqrt{0.45}\\
\sqrt{0.10}\\
\sqrt{0.45}
\end{array}
\right);\label{i1}\\
\left(
\begin{array}{l}
\zeta_{-1}(t=0)\\
\zeta_{0}(t=0)\\
\zeta_{+1}(t=0)
\end{array}
\right)
&=&
\left(
\begin{array}{l}
\sqrt{0.45}\imath\\
\sqrt{0.10}\\
\sqrt{0.45}
\end{array}
\right).\label{i2}
\end{eqnarray}
Namely, $90\%$ of the atoms are assumed to be initially prepared in the states with $m=\pm 1$. In Fig. \ref{fig6}(a, b) we show the time evolution of the populations of $^{87}$Rb and $^{23}$Na atoms in each spin state, i.e., the functions $|\xi_m(t)|^2$ and  $|\zeta_m(t)|^2$ ($m=0,\pm 1$). It is shown that each population rapidly oscillates with time with a small amplitude, around a slowly-varying central profile. The rapidly-oscillating details of these functions are essentially due to the nonlinearity of Eqs. (\ref{cp1})-(\ref{cp4}). In the following we will only focus on the behaviors of the slowly-varying central profiles which give coarse-grained descriptions for the spin dynamics.

Fig. \ref{fig6}(a, b) clearly shows that for each type of atom, the central profiles of populations 
$|\xi_{0}(t)|^2$ and  $|\zeta_{0}(t)|^2$ of the states with $m=0$ almost do not change with time. Thus the spin-changing processes $\langle 1,...,6\rangle$ defined in Eq. (\ref{sps}), in which the states with $m=0$ are involved, are all suppressed. On the other hand, Fig. \ref{fig6}(a, b) also show that the central profiles of  $|\xi_{\pm 1}(t)|^2$ and  $|\zeta_{\pm 1}(t)|^2$ significantly oscillate with time. This behavior implies that the singlet-pairing process $\langle 7\rangle$ is very apparent in our system. This result is shown more clearly in  Fig. \ref{fig6}(c), where we plot the relative magnetization
\begin{eqnarray}
M(t)=\left[|\xi_{+1}(t)|^2-|\xi_{-1}(t)|^2\right]+\left[|\zeta_{+1}(t)|^2-|\zeta_{-1}(t)|^2\right],\nonumber\\
\label{mt}
\end{eqnarray}
i.e., the population-difference of  the states before and after the process $\langle 7\rangle$, as a function of time.
It is shown that the central profile of $M(t)$ oscillates around zero with a significant amplitude. In Fig.~\ref{fig8}(a) we further show $M(t)$ for various value of  interaction intensity  $\gamma$. It is shown that, as in the two-body cases, the amplitude and period of the time oscillation of $M(t)$ clearly depends on the value of $\gamma$. Moreover, in our system the time evolution of $M(t)$ 
 also depends on the 
complex phase factors  of the initial state, i.e., ${\rm arg}[\xi_m(t=0)]$ and  ${\rm arg}[\zeta_m(t=0)]$ ($m=0,\pm 1$). It is clear that in our above calculations with initial condition (\ref{i1}, \ref{i2}), we have taken $\arg\left[\zeta_{-1}(t=0)\right]=0.5\pi$ and ${\rm arg}[\xi_{0,\pm 1}(t=0)]={\rm arg}[\zeta_{0,+1}(t=0)]=0$. As shown in Fig.  \ref{fig8} (b) these initial phase factors are modified, the behavior of $M(t)$ is also seriously changed.

{\color{black}For the single atomic species spinor BEC, the mean field  spin  dynamics  under  single-mode-approximation  has  been  mapped  to  one  nonrigid  pendulum \cite{PRA2005Zhang}.  The equal-energy contours there being described with two canonical parameters,  relative phase and the population fraction, shows that the spin dynamics is very sensitive to the initial states in some parameter regime.  For the two atomic species spinor BEC in our manuscript, the system can be mapped to three coupled nonrigid pendulums model \cite{Xu2012PRA}. The nonlinear behavior of spin dynamics can be understood by those pendulums models which are charactrized by three pairs of canonical conjugate variables. \\

}

\section{Conclusions and discussions}\label{secVI}
In this work we propose an approach for the enhancement and control of the singlet-pairing process between two  ultracold spin-1 atoms of different species, which  is based on the combination of the  real magnetic field and a  laser-induced specie-dependent  SMF. Taking the mixture of ultracold $^{87}$Rb and $^{23}$Na atoms as an example, we illustrate our approach for both a confined two-body system and a two-species spin-1 BEC. It is shown that the  singlet-pairing process can be enhanced while the other spin-changing process are suppressed, although the interaction intensity corresponding to the former one  is extremely weak. Therefore, our approach can be used for the observation of the singlet-pairing process and the precise measurement of the corresponding interaction parameter, as well as the entanglement generation of two different atoms.
Our method also can be applied to other atomic mixture systems in recently experiments, such as $^{7}$Li-$^{23}$Na mixture  \cite{Mil1128} and  $^{7}$Li-$^{87}$Rb mixture  \cite{fang2020collisional}.

\begin{acknowledgments}
This work is supported by the National Key R\&D Program of China (Grant No. 2018YFA0306502 (P.Z.)), NSFC Grants No.  U1930201(P.Z.) and No. 11674393 (P.Z.).
\end{acknowledgments}

\begin{widetext}
\appendix\section{Derivation of spin dynamic equations in Eq. (\ref{cp4})}\label{appendix1}
We consider the mixture of  BECs of spin-1 $^{87}$Rb BEC and $^{23}$Na BEC atoms with the real magnetic field and the laser  induced species-dependent SMF. The many-body Hamiltonian of our system is given by 
\cite{NaRbSL,Chen2018PRA}
\begin{eqnarray}\label{app_H}
\hat{H} & = & \hat{H}_{\rm Rb}+\hat{H}_{\rm Na}+\hat{H}_{\rm Rb-Na},\label{mixtureH}
\end{eqnarray}
where
\begin{eqnarray}
\nonumber \\
\hat{H}_{\rm Rb} & = & \int d\textbf{r}\Big[\hat{\Psi}_{m}^{\dag}\Big(-\frac{\hbar^{2}\nabla^{2}}{2M_{\rm Rb}}+V_{\rm Rb}({\bf r})+E_{m}^{\rm (Rb)}\Big)\hat{\Psi}_{m}+ \frac{\alpha_{\rm Rb}}{2}\hat{\Psi}_{i}^{\dag}\hat{\Psi}_{j}^{\dag}\hat{\Psi}_{j}\hat{\Psi}_{i}+\frac{\beta_{\rm Rb}}{2}\hat{\Psi}_{i}^{\dag}\hat{\Psi}_{k}^{\dag}F_{\nu,ij}F_{\nu,kl}\hat{\Psi}_{j}\hat{\Psi}_{l}\Big],\nonumber \\
\hat{H}_{\rm Na} & = & \int d\textbf{r}\Big[\hat{\Phi}_{m}^{\dag}\Big(-\frac{\hbar^{2}\nabla^{2}}{2M_{\rm Na}}+V_{\rm Na}({\bf r})+E_{m}^{\rm (Na)}\Big)\hat{\Phi}_{m}+ \frac{\alpha_{\rm Na}}{2}\hat{\Phi}_{i}^{\dag}\hat{\Phi}_{j}^{\dag}\hat{\Phi}_{j}\hat{\Phi}_{i}+\frac{\beta_{\rm Na}}{2}\hat{\Phi}_{i}^{\dag}\hat{\Phi}_{k}^{\dag}F_{\nu,ij}F_{\nu,kl}\hat{\Phi}_{j}\hat{\Phi}_{l}\Big],\nonumber \\
\hat{H}_{\rm Rb-Na} & = &\int d\textbf{r}\Big[\alpha\hat{\Psi}_{i}^{\dag}\hat{\Phi}_{j}^{\dag}\hat{\Phi}_{j}\hat{\Psi}_{i}+\beta\hat{\Psi}_{i}^{\dag}\hat{\Phi}_{k}^{\dag}F_{\nu,ij}F_{\nu,kl}\hat{\Phi}_{l}\hat{\Psi}_{j}+ \gamma\frac{(-1)^{i+j}}{3}\hat{\Psi}_{i}^{\dag}\hat{\Phi}_{-i}^{\dag}\hat{\Phi}_{-j}\hat{\Psi}_{j}\Big].
\end{eqnarray}
Here the repeated subscripts means summations for $i,j,k,l,m=\pm1,0$ and $\nu=x,y,z$.
 $\hat{\Psi}_{m}(\textbf{r})$ and $\hat{\Phi}_{m}(\textbf{r})$ ($m=0,\pm 1$)
 are the annihilation operators of $^{87}{\rm Rb}$ and $^{23}{\rm Na}$ atom with magnetic quantum number $m$ at position ${\bf r}$. $V_{\rm Rb}({\bf r})$ and $V_{\rm Na}({\bf r})$ are the trapping potentials for $^{87}$Rb  and $^{23}$Na atoms, respectively. The interaction strength coefficients $\alpha_{\rm Na,Rb},~\beta_{\rm Na,Rb},~\alpha,~\beta$ and $\gamma$ are defined in Eqs. (\ref{lj}-\ref{gamma}), and $F_{\nu=x,y,z}$ are the spin-1 matrices
\begin{equation}
F_{x}=\frac{1}{\sqrt{2}}\left(\begin{array}{ccc}
0 & 1 &0\\
1 &0 &1\\
0 &1 & 0
\end{array}\right),\ F_{y}=\frac{1}{\sqrt{2}}\left(\begin{array}{ccc}
0 & -\imath &0\\
\imath &0 &-\imath\\
0 &\imath & 0
\end{array}\right),\ F_{z}=\frac{1}{\sqrt{2}}\left(\begin{array}{ccc}
1 & 0 &0\\
0 &0 &0\\
0 &0 & -1
\end{array}\right).
\end{equation}

Now we apply the mean-field approximation for our system, under which each atom of the same species is in the same one-body state. Furthermore, since in our system the spin-independent interaction intensities (i.e., the $\alpha$-parameters) are much stronger than the spin-dependent interaction intensities (i.e., the $\beta$-parameters and $\gamma$), we further use single-mode approximation (except some dynamical mean-field induced resonant regimes \cite{Jie1,Jie2}) under which the spatial wave function of each atom is spin-independent, and is only determined by the spin-independent interaction. In our calculations based on the above approximations, 
each $^{87}{\rm Rb}$ atom is in the same state corresponding to the wave function
  \begin{eqnarray}
 \psi({\bf r})\left[\sum_{m=0,\pm 1}\xi_m(t)|m\rangle_{\rm Rb}\right],\label{a4}
 \end{eqnarray}
and each $^{23}{\rm Na}$ atom is in the same state corresponding to the wave function
  \begin{eqnarray}
 \phi({\bf r})\left[\sum_{m=0,\pm 1}\zeta_m(t)|m\rangle_{\rm Na}\right],\label{a5}
 \end{eqnarray}
with 
\begin{eqnarray}
\int d{\bf r}|\psi({\bf r})|^2=\int d{\bf r}|\phi({\bf r})|^2=\sum_{m}|\xi_{m}|^{2}=\sum_{m}|\zeta_{m}|^{2}=1.\label{nnor}
\end{eqnarray}
 Here the spatial wave functions $\psi({\bf r})$ and $\phi({\bf r})$ are determined by the 
 Gross-Pitaevskii equations for the system without the spin-independent interactions:
 \begin{eqnarray}
~\left[-\frac{\hbar^{2}\nabla^{2}}{2M_{\rm Rb}}+V_{\rm Rb}({\bf r})+\alpha_{\rm Rb}N_{\rm Rb}|\psi(\textbf{r})|^{2}+\alpha N_{\rm Na}|\phi(\textbf{r})|^{2}\right]\psi(\textbf{r}) & = & \mu_{\rm Rb}\psi(\textbf{r}), \label{cgpmu1}\\
~\left[-\frac{\hbar^{2}\nabla^{2}}{2M_{\rm Na}}+V_{\rm Na}({\bf r})+\alpha_{\rm Na}N_{\rm Na}|\phi(\textbf{r})|^{2}+\alpha N_{\rm Rb}|\psi(\textbf{r})|^{2}\right]\phi(\textbf{r}) & = & \mu_{\rm Na}\phi(\textbf{r}),\label{cgpmu}
\end{eqnarray}
where $N_{\rm Rb}$ and $N_{\rm Na}$ are the numbers of $^{87}$Rb  and $^{23}$Na atoms, respectively, and
$\mu_{\rm Rb}$ and $\mu_{\rm Na}$ are the corresponding chemical potentials.
It is clear that the wave functions (\ref{a4}, \ref{a5}) are just the ones in Eq. (\ref{wff}) of Sec. V. 

Now we derive the dynamical equation for the coefficients $\xi_m(t)$ and $\zeta_m(t)$ ($m=0,\pm 1$). Under the above mean-field and single-mode approximations, the time-dependent many-body state $|{\rm Q}(t)\rangle$ of our system can be expressed as 
\begin{eqnarray}
|{\rm Q}(t)\rangle=\frac{1}{\sqrt{N_{\rm Rb}!}}\left(\sum_{m=\pm1,0}
\xi_{m}\int d{\bf r}_{\rm Rb}\psi^{\ast}({\bf r}_{\rm Rb}){\hat \Psi}^\dagger_m({\bf r}_{\rm Rb})
\right)^{N_{\rm Rb}}\otimes
\frac{1}{\sqrt{N_{\rm Na}!}}
\left(\sum_{m=\pm1,0}\zeta_{m}\int d{\bf r}_{\rm Na}\phi^{\ast}({\bf r}_{\rm Na}){\hat \Phi}^\dagger_m({\bf r}_{\rm Na})
\right)^{N_{\rm Na}}|{\rm vac}\rangle,\nonumber\\
\end{eqnarray}
with $|{\rm vac}\rangle$ being the vacuum state. Furthermore, the instantaneous average energy of our system on this many-body sate can be expressed as a function of the coefficients $\xi_m(t)$ and $\zeta_m(t)$, i.e., 
 \begin{eqnarray}
E_{\rm MB}[\xi_{0,\pm 1}(t);\xi_{0,\pm 1}^{\ast}(t);\zeta_{0,\pm 1}(t);\zeta_{0,\pm 1}^{\ast}(t)]\equiv \langle {\rm Q}(t)|{\hat H}|{\rm Q}(t)\rangle.
\end{eqnarray}
Thus, using the time-dependent variational principle, we can obtain the dynamical equations for $\xi_m(t)$ and $\zeta_m(t)$ ($m=0,\pm 1$) (up to a global
phase factor):
\begin{eqnarray}
\imath\frac{d}{dt}\xi_m(t)=\frac{\partial E_{\rm MB}}{\partial \xi_m^\ast};\label{iie1}\\
 \imath\frac{d}{dt}\zeta_m(t)=\frac{\partial E_{\rm MB}}{\partial \zeta_m^\ast}.\label{iie2}
\end{eqnarray}
With straightforward calculations, one can find that Eqs. (\ref{iie1}, \ref{iie2}) for $m=0,\pm 1$ are just  Eqs. (\ref{cp1}-\ref{cp4}) in our maintext (up to a global phase factor).

  \section{Calculation of atomic probability densities via Thomas-Fermi approximation}\label{appendix2}
  
 In this appendix we derive the atomic  probability densities $|\psi({\bf r})|^2$ and $|\phi({\bf r})|^2$ via Thomas-Fermi approximation \cite{pethick_smith_2008,Wang_2015}. Under this approximation, the Gross-Pitaevskii equations in Eq. (\ref{cgpmu1}, \ref{cgpmu}) can be simplified as
 \begin{eqnarray}
\left[V_{\rm Rb}({\bf r})+\alpha_{\rm Rb}|\psi({\bf r})|^2+\alpha |\phi({\bf r})|^2\right]\psi(\textbf{r}) & = & \mu_{\rm Rb}\psi(\textbf{r}), \\
\left[V_{\rm Na}({\bf r})+\alpha_{\rm Na}|\phi({\bf r})|^2+\alpha |\psi({\bf r})|^2\right]\phi(\textbf{r}) & = & \mu_{\rm Na}\phi(\textbf{r}).
\end{eqnarray}
Here we assume the atoms are in the isotropic harmonic trap with the frequency $\omega_{\rm Rb}$ for $^{87}$Rb BEC and $\omega_{\rm Na}$ for $^{23}$Na BEC, thus the trap potentials are $V_{\rm Rb,Na}=M_{\rm Rb,Na}\omega_{\rm Rb,Na}^{2}r^{2}/2$.
The solutions of these equations are 
\begin{eqnarray}
|\psi(\textbf{r})|^{2}&=&\frac{2\mu_{\rm Rb}X_{\rm Rb}-\omega_{\rm Rb}^{2}r^{2} Y_{\rm Rb}}{2N_{\rm Rb}Z}\Theta\left(r-r_{\rm tf,Rb}\right),\label{den1} \\
|\phi(\textbf{r})|^{2}&=&\frac{2\mu_{\rm Rb}X_{\rm Na}-\omega_{\rm Rb}^{2}r^{2} Y_{\rm Na}}{2N_{\rm Na}Z}\Theta\left(r-r_{\rm tf,Na}\right), \label{den2}
\end{eqnarray}
where $\Theta(z)$ is the step function which satisfies $\Theta(z)=1$ for $z>0$ and $\Theta(z)=0$ for $z\leq 0$, $r_{\rm tf,Rb}$ and $r_{\rm tf, Na}$ are the Tomas-Fermi radius,
\begin{eqnarray}
r_{\rm tf,Rb} = \frac{\sqrt{2\mu_{\rm Rb}}}{\omega_{\rm Rb}}\sqrt{\frac{X_{\rm Rb}}{Y_{\rm Rb}}};\ \ {\color{black}r_{\rm tf,Na} = \frac{\lambda_{\omega}}{\sqrt{\lambda_{\mu}}} \frac{\sqrt{2\mu_{\rm Na}}}{\omega_{\rm Na}}\sqrt{\frac{X_{\rm Na}}{Y_{\rm Na}}}.}~~\label{tf}
\end{eqnarray}
The coefficients $X_{\rm Rb,Na}$ are related to the chemical potential ratio $\lambda_{\mu}=\mu_{\rm Na}/\mu_{\rm Rb}$,
\begin{eqnarray}
X_{\rm Rb}=\lambda_{\mu}\alpha -\alpha_{\rm Na};~~ X_{\rm Na}=\alpha-\lambda_{\mu}\alpha_{\rm Rb},~~ 
\end{eqnarray}
and $Y_{\rm Rb,Na}$ are related to the trap frequency ratio $\lambda_{\omega}=\omega_{\rm Na}/\omega_{\rm Rb}$,
\begin{eqnarray}
Y_{\rm Rb}=\lambda_{\omega}^{2}\alpha M_{\rm Na}-\alpha_{\rm Na} M_{\rm Rb};~~Y_{\rm Na}=\alpha M_{\rm Rb}-\lambda_{\omega}^{2}\alpha_{\rm Rb} M_{\rm Na}.
\end{eqnarray}
$Z$, $Z=\alpha^2-\alpha_{\rm Rb}\alpha_{\rm Na}$, is a positive constant coefficient in this system and the chemical potentials $\mu_{\rm Rb, Na}$ are determined by the normalization condition (\ref{nnor}).


It is clear that both the probability densities $|\psi(\textbf{r})|^{2}$ and $|\phi(\textbf{r})|^{2}$ and the Tomas-Fermi radius $r_{\rm tf,Rb}$ and $r_{\rm tf,Na}$ should be positive. This yields that the Thomas-Fermi approximation can be used for the systems with
$1.998<\omega_{\rm Na}/\omega_{\rm Rb}<2.658$ and $1.056<\lambda_{\mu}<1.869$.
\end{widetext}
\bibliography{Citations}   

\begin{thebibliography}{60}
\bibitem [{res()}]{resonant_conditions}%
  \BibitemOpen
  \href@noop {} {\bibinfo  {journal} {{\color{black}The exact value of
  $g_{\left\langle l\right\rangle}$ ($l=1,...,7$) is determined by the detail
  of the systems. For instance, for the two-atom system of Sec. IV the relevant
  interaction intensity $g_{\left\langle 3,4,7\right\rangle}$ are the
  non-diagonal elements of the matrix of Eq. (\ref{calm}), with the expressions
  being given by Eqs. (\ref{ge1}-\ref{ge3})}}\ }\BibitemShut {NoStop}%
\bibitem [{tra({\natexlab{a}})}]{trap1}%
  \BibitemOpen
  \href@noop {} {\bibfield  {journal} {\bibinfo  {journal} {This conclusion can
  be obtained via the following analysis. The effect of the transition between
  the ground state and the $n$-th excited states of $H_{{\rm ho}}$ can be
  estimated via the ratio $|D/(\varepsilon_{n}-\varepsilon_{0})|$, where
  $\varepsilon_{n}$ and $\varepsilon_{0}$ are the eigen-energies of $H_{{\rm
  ho}}$ corresponding to these two states and $D$ is the corresponding matrix
  elment of $U_{{\rm Rb-Na}}$. Direct calculation shows that
  $|D/(\varepsilon_{n}-\varepsilon_{0})|\lesssim|a_{{\rm
  Rb-Na}}^{(0,1,2)}/a_{{\rm ho}}|$. Thus, when the scattering lengths are much
  smaller than $a_{{\rm ho}}$ we have
  $|D/(\varepsilon_{n}-\varepsilon_{0})|\ll1$ which yields the transition can
  be neglected.}\ } ({\natexlab{a}})}\BibitemShut {NoStop}%
\bibitem [{tra({\natexlab{b}})}]{trap2}%
  \BibitemOpen
  \href@noop {} {\bibfield  {journal} {\bibinfo  {journal} {In this
  measurement, if it is required, one can go beyond the approximation that the
  atomic spatial motion are frozen in the state $|0\rangle_r$, and derive the
  exact solution for the two-body problem with the approach in e.g., the ones
  in [busch, calarco, chenyue, sala], and compare the exact solution with
  experimental measurements}\ } ({\natexlab{b}})}\BibitemShut {NoStop}%
\end{thebibliography}
\end{document}